# Correlation between the rheology of 2D-inks precursors and the droplet size generated from a capillary nozzle in dripping regime

*Pedro C. Rijo, Josué M. O. Cremonezzi, Ricardo J. E. Andrade, and Francisco J. Galindo-Rosales\**

Pedro C. Rijo

Transport Phenomena Research Center (CEFT), Mechanical Engineering Department, Faculty of Engineering, University of Porto, 4200-465 Porto, Portugal.

ALiCE—Associate Laboratory in Chemical Engineering, Faculty of Engineering, University of Porto, Rua Dr. Roberto Frias, 4200-465 Porto, Portugal

Josué M. O. Cremonezzi

School of Engineering, Mackenzie Presbyterian University, Rua da Consolação 930, São Paulo - SP. 01302-907, Brazil

MackGraphe - Mackenzie Institute for Research in Graphene and Nanotechnologies, Mackenzie Presbyterian Institute, Rua da Consolação, 896 São Paulo – SP, 01302-907, Brazil

Ricardo J. E. Andrade

School of Engineering, Mackenzie Presbyterian University, Rua da Consolação 930, São Paulo - SP. 01302-907, Brazil

MackGraphe - Mackenzie Institute for Research in Graphene and Nanotechnologies, Mackenzie Presbyterian Institute, Rua da Consolação, 896 São Paulo – SP, 01302-907, Brazil

Francisco J. Galindo-Rosales

Transport Phenomena Research Center (CEFT), Chemical Engineering Department, Faculty of Engineering, University of Porto, 4200-465 Porto, Portugal.

ALiCE—Associate Laboratory in Chemical Engineering, Faculty of Engineering, University of Porto, Rua Dr. Roberto Frias, 4200-465 Porto, Portugal

E-mail: galindo@fe.up.pt

**Abstract:** This study provides a complete rheological characterization of 2D nanomaterial dispersions, employed as 2D-inks precursors in printed electronics. Three different 2D nanomaterials (molybdenum disulfide ($MoS_2$), graphene, and hexagonal boron nitride(hBN)) were dispersed in a Newtonian fluid (toluene) and a viscoelastic fluid (toluene + ethyl cellulose) with different polymer concentrations. The presence of nanoparticles does not change the shear rheology of the carrier fluid. Regarding the extensional rheology, the results showed that the pinch-off phenomenon is present in all Toluene suspensions; however, the presence of the ethyl cellulose introduces elasticity in the system, even leading to the formation of beads-on-a-string, and the relaxation times of the suspensions depends on the kind of nanoparticles present in the fluid. As controlling the droplet size when dispensing 2D-inks is of paramount importance for printed electronics, as well as for many other applications, here it is presented a correlation between the rheological properties of these 2D-inks precursors and their droplet size when generated from a capillary nozzle in dripping regime.

## 1. Introduction

In the last decade, the world of consumer electronics has experienced massive improvements in the manufacturing techniques toward the production of smaller, thinner, and flexible products for everyday use. Whereas the use of traditional solid-state technology shows some limitations to the production of these types of devices, different printing techniques, such as Ink-jet printing, screen printing and gravure printing, offer a low-cost, simple, and scalable production methods to respond to these new challenges.[1] These technologies require functional inks, which have markedly different chemical, physical and rheological properties depending on the size, shape, quality of the dispersion, concentration, and type of filler present, as well as on the carrier fluid.[2]

Graphene, hexagonal boron nitride (hBN) and molybdenum disulfide ($MoS_2$) inks have recently been applied in the production of electronic circuits, sensors and coating films for batteries[3] using different printing techniques.[1, 4] The selection of the most suitable printing process for each ink depends on the rheological properties of the ink; moreover, the technical conditions of the chosen printing process can be tweaked to provide the best resolution.[5] Most works published in this area have only briefly studied the rheological properties of these inks, mainly in the shear rheology, forgetting the extensional rheology; however, the characterization under extensional flow is of paramount importance to the efficient operation of different printing and coating processes where the extensional flow dominates a portion of the fluid domain, such as in the generation of droplets.[6]

The simplest way to create single droplets is by a slow ejection of fluid through a capillary tube, which allows the formation of a pending drop of fluid attached to the perimeter of the tube by means of the surface tension. According to Rayleigh formula, once the droplet achieves a critical size ($m \approx \frac{3.8\sigma r}{g}$ being $m$ the mass, $\sigma$ the surface tension and $r$ the radius of the capillary tube), it is released.[7] This mechanism implies low production rates and large droplet sizes, controlled by the shape and the size of the capillary tube. Thus, for Newtonian fluid in dripping regime, the droplet size is determined by the density and surface tension of the fluid and gravity.[8] However, the drop size is around 10% smaller for a viscoelastic liquid.[9] The presence of micron-sized spherical particles at low concentrations in a Newtonian fluid does not affect the dispensing characteristics, compared to those of the Newtonian carrier fluid, but strongly affects the drop size of elastic liquids.[9] Despite the previous studies and the importance of 2D nanomaterials in printed electronics applications, to the best of authors' knowledge, it has not been yet studied how the presence of 2D nanoparticles affects the dispensing characteristics of Newtonian and viscoelastic fluids in the dripping regime.

To investigate the role of 2D nanoparticles on the dripping dynamics of Newtonian and viscoelastic fluids, it is important to understand their rheological properties. In this study we use graphene, hexagonal boron nitride, and molybdenum disulfide as 2D nanomaterials.. The structures of these nanomaterials promote mechanical robustness, high thermal and electrical conductivity, and chemical inertness.[10] The weak van der Waals forces between the layers facilitate the exfoliation during the formulation of electronic inks.

The steady-shear viscosity of suspensions is strongly affected by the nanoparticles volume fraction, as well as their shape and spatial arrangements.[11] Timofeeva *et al.* reported that elongated particles increase the shear viscosity of nanofluids rather than spherical nanoparticles.[12] So, the presence of non-spherical nanoparticles increases the energy dissipation due to the particle-fluid interactions and particle-particle interactions.[11] Moreover, another study verified that the increment viscosity is strongly affected by the high aspect ratio of the nanoparticles.[13] This happens because the suspension viscosity is higher when elongated and plate-like rather than spherical nanoparticles are present as a result of the higher contact area and the consequent higher surface interaction between the particles and the carrier fluid. when the particles density and concentration are the same.[13]

Marra *et al.*[4c] studied the application of graphene nanoplatelet (GNP)-based ink to produce smart textile strain sensors via screen printing technique. The authors performed a shear rheological characterization of the GNP ink and verified that a shear-thinning behavior at shear rates exceeding 100 s$^{-1}$. They found that the rapid decrease of suspension viscosity occurs due to the lubricant effect of GNPs particles. Further, Moraes *et al.*[4e] studied the application of hBN nanosheets inks for printed electronics and electronic coating through shear rheological characterization. The authors found that the steady-shear viscosity increases when the concentration of hBN nanoparticles increases for the same shear rate. They also observed that the presence of hBN induced the shear-thinning behavior. Moreover, Moraes *et al.* verified the addition of ethyl cellulose (EC) promotes the stability of the ink.

In addition, Wang *et al.* studied the application of graphene/molybdenum disulfide inks with several concentrations of particles to apply in supercapacitors production performing a shear rheological characterization.[4d]. They found that the viscosity for all inks decreased significantly when the shear rates increased (shear-thinning behavior). Furthermore, the authors found that shear-thinning fluids benefit from the smooth transfer of the ink during the printing process without the formation of defects.

In the present work, the attention is focused on the assessment of the rheological properties of precursors of 2D inks under shear and uniaxial extensional flows. Furthermore, the chemical-physical properties for 2D nanoparticles suspensions and the dependency between the concentration of polymer and the appearance of beads-on-a-string (BOAS) during the filament breakup is highlighted. Finally, a dimensionless analysis is performed in order to determine an experimental correlation for the droplet size as a function of the main dimensionless parameters controlling the dripping mode regime.[14]

## 2. Materials and Methods

### 2.1. Materials

The reagents used for the formulation of 2D nanoparticles suspensions were ethyl cellulose (48% ethoxyl basis), toluene (purity > 99.9%), graphene nanoplatelets (GNP), hBN powder, and $MoS_2$ powder. All material was supplied by Sigma-Aldrich, except toluene, supplied by Carlos Erba Reagents, and GNP, which was graciously provided by Graphenest.

### 2.2. Formulation of 2D nanoparticle Inks

The preparation of the 2D nanoparticle suspensions was defined based on the protocols reported by Moraes *et al.*[4e, 15] The procedure consisted on the following consecutive steps: (i) dissolve ethyl cellulose in toluene with different concentrations (0%, 2.5%, and 5% w/v) using a magnetic stirrer for 24 hours at 20 °C; (ii) add the powder of GNP or hBN or $MoS_2$ with a concentration of 0.2 mg mL$^{-1}$; and (iii) dispersion and exfoliation of nanoparticles are made through ultrasonic bath for 3 hours, where the temperature of the bath cannot exceed 50 °C.

Before the chemical, physical and rheological properties of suspensions determination, it is important to guarantee their stability during the experimental tests. So, a rudimentary sedimentation study was performed by naked eye watching 5 mL samples of at rest for 1 hour. When the carrier fluid was pure toluene, suspensions were stable for 5 minutes and easily re-dispersible by stirring. For suspensions in which the carrier fluid was the polymeric solution, suspensions were stable for at least 1 hour. Thus, rheological tests duration was chosen as 5 min, time in which was possible to ensure homogeneity condition based on the 1-hour sedimentation test.

## 2.3 Characterization of GNP, hBN and $MoS_2$ Nanoparticles

Before samples' preparation, the fluids were sonicated for 10 min to ensure homogenization. Dynamic light scattering (DLS) was conducted in a Litesizer 500 particle analyzer (Anton Paar). Analyses were conducted after 1 min equilibration time at 25 °C, in 60 runs of 10 s. Fourier transform infrared (FTIR) spectroscopy was performed in an IRAffinity-1 infrared spectrophotometer (Shimadzu). Spectra were obtained after 32 accumulations in 4 $cm^{-1}$ resolution in the attenuated total reflection (ATR) mode. For Raman spectroscopy, drops of each fluid were deposited in silicon wafers. The samples were dried in an oven overnight at 40 °C. Measurements were done in an Alpha 300R spectrometer (WITec) using a laser with a wavelength of 532 nm.

## 2.4. Rheological characterization of the 2D nanoparticle suspensions

### 2.4.1. Shear Rheology

To determine the viscosity curves of the suspensions, a controlled shear stress Anton Paar MCR301 rheometer equipped with a plate-plate geometry of 50 mm diameter and a gap distance of 0.1 mm was used. All tests were performed at 20 °C. For each suspension, 5 independent runs were done to ensure a good repeatability of the results.

There are two limitations that if not considered can lead to misinterpretation of the results obtained.[16] The first limitation is the minimum torque at which all points measured with a higher torque are credible. In this work, the minimum torque is set at 1.1 μN.m. The second limitation is the presence of secondary flows at high shear rates. According to Ewoldt et al.[16], the sample inertia can cause nonideal velocity fields during the steady flow, which can cause secondary flows superposed on the primary simple shear flow due to finite volume of fluid and curved streamlines in unstable configurations. Based on these two limitations, steady-shear viscosity tests were performed for a range of shear rates of 1 to $10^5$ s$^{-1}$, which allowed to ensure being above the minimum torque limit and below the onset of secondary flows.

*2.4.2. Extensional Rheology*

A capillary breakup extensional rheometer (CaBER) was used to perform the extensional rheological characterization. CaBER imposes an extensional step strain of order unity, and the unbalanced capillary forces are responsible for the filament thinning process. Far from the end-plates, the fluid filament undergoes an uniaxial extensional flow in which the extension rate is dictated by the surface tension of the fluid and its extensional properties.[17] The CaBER only records the time evolution of the midpoint diameter of the thinning filament and the relaxation time for viscoelastic fluids is calculated from the exponential decay of the minimum filament diameter with time.[17] Ng and Poole[18] highlighted the importance of using high-speed imaging coupled with CaBER. They showed that errors induced on average over the finite thickness of a laser micrometer can be up to 40% in the case of low-viscous Newtonian fluids. Further, the authors proved that for a deformation thinning power-law fluid the impact of laser micrometer misalignment with the location of filament breakup can lead to significant errors in the apparent extensional viscosity. Both errors induced by these two types of fluids are due to filament curvature.[18] In addition, the filament evolution of many yield stress fluids exhibits conical tapering prior to breakup where the axial location of the breakup is typically not known a priori. Based on this evidence, the extensional characterization of the fluids proposed in this work was done using the high-speed imaging system coupled to the CaBER. The experimental setup for the extensional rheometry is depicted in **Figure 1**, which allowed to successfully characterize other particulate suspensions in previous studies.[5, 19] The evolution of the filament diameter was recorded using a high-speed camera (Photron FASTCAM mini UX100) coupled with a set of optical lenses (Optem Zoom 70XL) with a variable magnification from 1X to 5.5X. To visualize the filament shape, a 60 mm Telecentric Backlight Iluminator (TECHSPEC) in which a white light was supplied by a fiber optic cable connected to a light source (Leica EL6000) was employed. The image analysis was done using an in-house developed code in MATLAB, with which the time evolution of the minimum filament radius is determined.

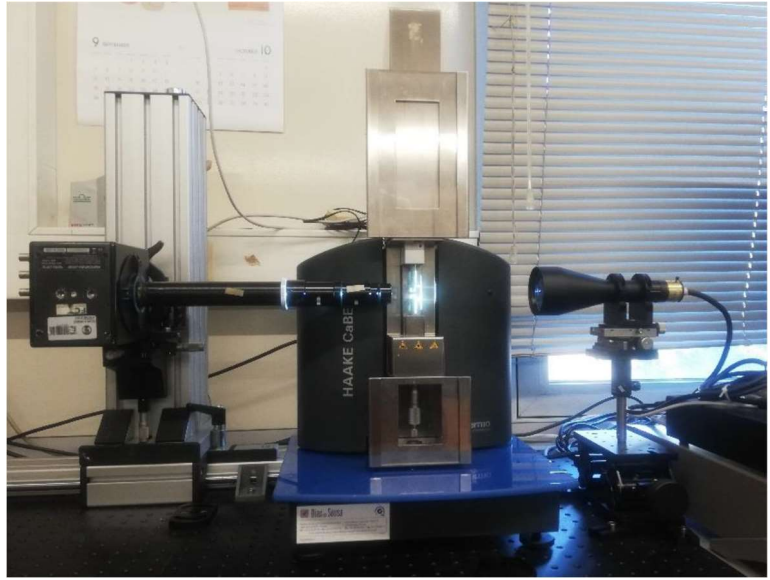

**Figure 1.** Experimental setup used to measure the time evolution of the minimum filament radius.

According to Rodd *et al.*[20], the minimum relaxation time measurable in CaBER is the order of 1 ms. Moreover, relaxation times of 1 ms are difficult to measure, especially for low viscosity fluids or for fluids with low surface tension.[21] This happens due to inertial effects and the short time for the thinning filament and break up. To overcome these challenges, Campo-Deaño and Clasen[22] developed the slow retraction method (SRM). In the SRM method, the filament thinning is promoted by a slow extension of the liquid bridge, i.e., low extensional speed of the upper plate, opposite to the fast step strain preset in CaBER. Using this method, inertia effects are minimized and relaxation times lower than 240 µs can be measured.[22]

## 2.5. Droplet Study

The experimental setup used to study the droplet sizes and the dripping-jetting transition for the 2D nanoparticle suspensions is shown in **Figure 2.** In this experimental setup a Centoni Nemesys S syringe pump equipped with a 2.5 mL Hamilton gas tight syringe was used to inject fluid in vertical direction into stagnant air. The dispensing and droplet formation process was recorded using the same optical setup used for the extensional rheometry, i.e., a high-speed camera (Photron FASTCAM mini UX100) coupled with a set of optical lenses (Optem Zoom 70XL) with a variable magnification from 1X to 5.5X, retro-illuminated with white light supplied by a fiber optic cable connected to a light source (Leica EL6000) and a 60 mm Telecentric Backlight Illuminator (TECHSPEC).

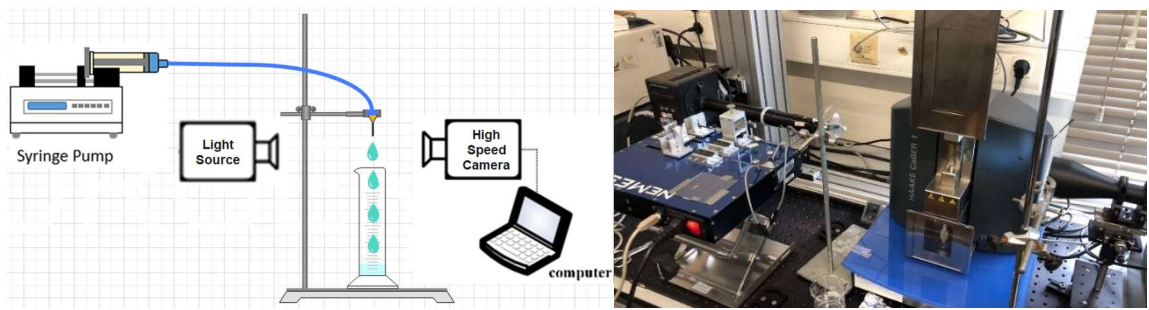

**Figure 2.** Schematic diagram of the experimental setup used to study the dispensing regime (left). Picture of the experimental setup (right)

**Table 1.** Geometrical dimensions of the needle used in this work.[23]

| Needle Color | Outside Diameter [mm] | Inner Diameter [mm] |
|---|---|---|
| Amber | 1.65 | 1.36 |
| Blue | 0.72 | 0.41 |
| Orange | 0.65 | 0.33 |
| Pink | 0.91 | 0.61 |

In this work, the influence of the nozzle diameters on the droplet size was studied (**Table 1**). The nozzles consisted of general-purpose-stainless-steel tips provided by Nordson Corporation having flat edges, trying to approximate the real conditions used to produce drops through printing processes, such as inkjet printing.

### 3. Results and Discussion

### 3.1. Structural Characterization of GNP, hBN and $MoS_2$

*3.1.1. Dynamic Light Scattering (DLS)*

The size of the particles was evaluated by DLS. **Figure 3a** shows the relative frequency of the nanomaterials' particle size. Compared to the other nanomaterials, it was found that GNP presented the biggest range of particle sizes, varying from 35 to 635 nm. Despite the wider distribution, GNP had particles with a smaller hydrodynamic diameter, equal to 92.7 nm. Nevertheless, the nanomaterial showed a $D_{90}$ of 261.0 nm, i.e., 90% of the material's particles had sizes up to such a size, as shown in **Figure 3b**.

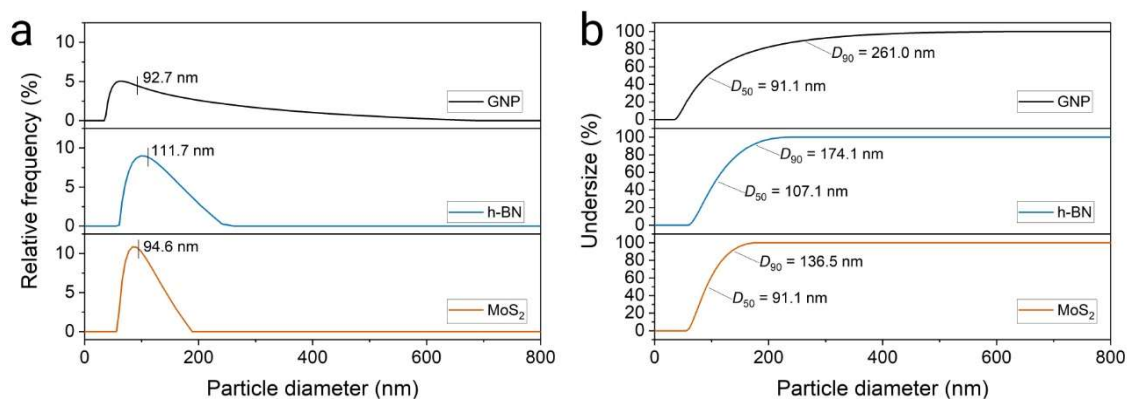

**Figure 3. (a)** Relative frequency and **(b)** undersize DLS curves of the nanomaterials.

hBN showed sizes between 60 and 240 nm, with a hydrodynamic diameter equal to 111.7 nm, and 90% of the material's particles had sizes up to 174.1 nm. MoS$_2$ particles' sizes presented the narrowest distribution, ranging from 60 to 175 nm. The hydrodynamic diameter of 94.6 nm was close to that of GNP, however, MoS$_2$ D$_{90}$ was much smaller, of only 136.5 nm. To simplify the data visualization from **Figure 3, Table 2** shows the maximum size of 90% of particles dispersed in toluene (D$_{90}$) and also the hydrodynamic diameter (D$_H$).

**Table 2**. Particle size measured from DLS technique for GNP, hBN, and MoS$_2$ dispersed in toluene.

| Particles | D$_{90}$ [nm] | D$_H$ [nm] |
|---|---|---|
| **GNP** | 261.0 | 92.7 |
| **hBN** | 174.1 | 111.7 |
| **MoS$_2$** | 136.5 | 94.6 |

*3.1.2. Fourier transform infrared spectroscopy (FTIR)*

The main bands observed in the suspensions infrared spectra could be assigned to toluene vibration modes.[24] As shown in **Figure 4a**, the modes between 3200-2960 cm$^{-1}$ were related to the C–H bonds stretching vibration (ν). From 2960 to 2790 cm$^{-1}$, bands were assigned to the ν of the methyl C–H bonds. In 1608 and 1028 cm$^{-1}$ the ν of C–C modes appeared. Bending deformation (δ) modes of C–H bonds from the methyl and ring gave rise to the bands at 1489 and 1085 cm$^{-1}$, respectively. The most prominent bands were the C–H wagging (w) at 724 cm$^{-1}$ and the one related to the ring torsion at 690 cm$^{-1}$. There were no relevant differences between the pure toluene and the GNP/toluene dispersion spectra. Nevertheless, the fluids with EC showed increased absorbance of the ν C–H between 3200 and 2790 cm$^{-1}$ and additional bands in the 1200-950 cm$^{-1}$ region. This region, plotted in **Figure 4b** for better visualization, comprehends bands related to the ν of C–O–C bonds in EC.[25]

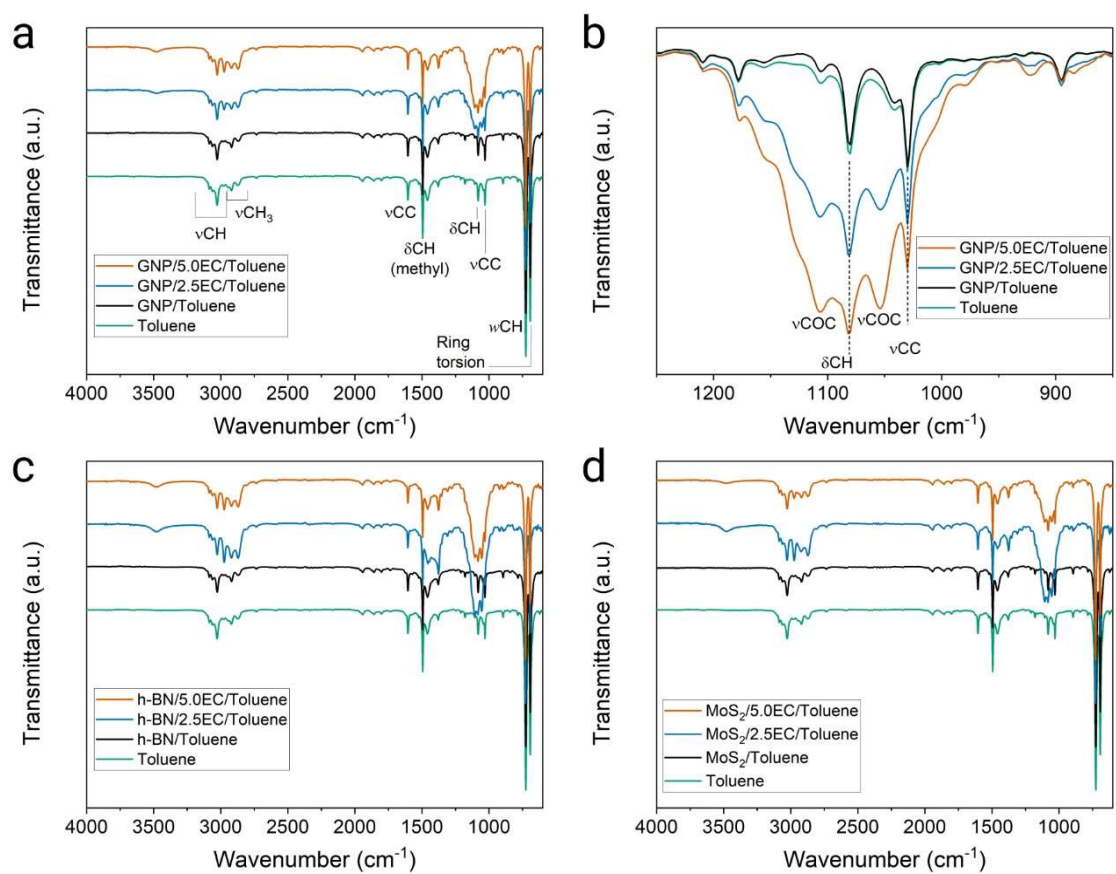

**Figure 4.** Infrared spectra of **(a)** GNP/, **(b)** hBN/, and **(c)** MoS$_2$/ethyl cellulose/toluene model fluids.

As well as GNP/toluene spectrum was similar to the toluene spectrum, were hBN/toluene and MoS$_2$/toluene spectra. This is because, compared to toluene and EC, the GNP[26], MoS$_2$[27], and hBN[28] infrared absorption is too low, due to distinct levels of polarization. Besides, even if the bands were strong enough, the superposition of the toluene bands would mask them.

*3.1.3. Raman spectroscopy*

Fluids had to be dried for Raman spectroscopy. Therefore, toluene bands[29] were not observed in the Raman spectra of the nanomaterials/EC mixtures. **Figure 5a** shows the Raman spectra of GNP and GNP/EC mixtures. GNP exhibited the typical graphitic modes, i.e., the D, G, and 2D bands at 1347, 1575, and 2700 cm$^{-1}$.[30] The D and G bands intensity ratio (ID/IG) of 0.12 showed that GNP had good structural integrity.[30b] On the other hand, celluloses exhibit band-rich Raman spectra, with the bands' position varying depending on the cellulose origin. The main bands on the cellulose Raman spectrum are those related to the ν vibration of the $CH_2$ ring bonds at ~2970, 2935 and 2890 $cm^{-1}$. Several bands regarding the bending deformation of these bonds appear in the 1280-1480 $cm^{-1}$ range.[31] Bands at ~250, 895, 1120, and 3240-3290 $cm^{-1}$ may be assigned to the twisting (τ) vibration of the C–OH, symmetric in-plane ν C–O–C, symmetric glycosidic ν C–O–C, and free ν OH, respectively.[31b] These were the bands observed in the spectra of the nanomaterials/EC mixtures, superposed to the GNP bands. Moreover, the cellulose bands' relative intensity increased when the EC proportion was doubled.

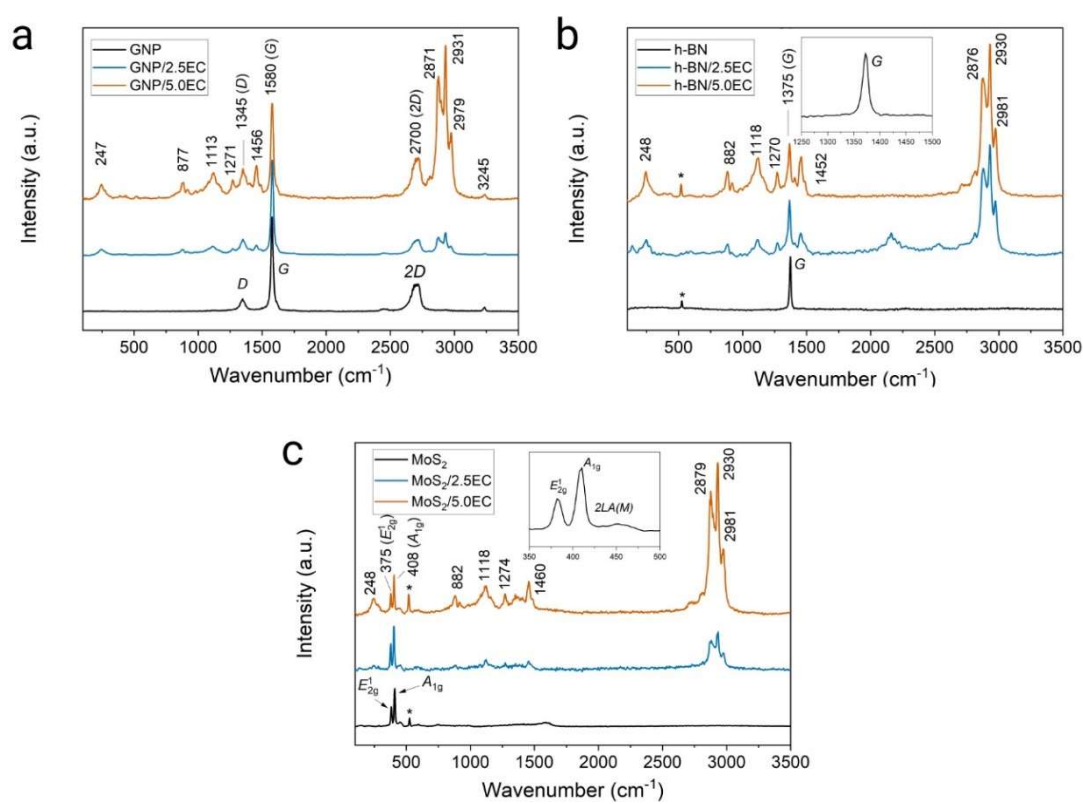

**Figure 5.** Raman spectra of **(a)** GNP/, **(b)** hBN/, and **(c)** MoS$_2$/ethyl cellulose. Insets in **(b)** and **(c)** show the interest region of the pure nanomaterial's spectra. Bands marked with an asterisk (*) are assigned to the silicon substrate.

Similar conclusions could be obtained from the analysis of the hBN/EC and MoS$_2$/EC Raman spectra (**Figure 5b** and **c**). Pure hBN presented the characteristic G band at 1375 $cm^{-1}$, attributed to $E_{2g}$ phonon mode at the Brillouin zone center.[28, 32]. MoS$_2$, in turn, showed the two typical modes, namely, the $E_{2g}^1$, which occurs due to the opposite in the plane vibration of 2 atoms of S concerning a Mo atom, and the $A_{1g}$, resulting of out-of-plane vibration of S atoms in opposite directions.[33] Such modes appeared in 382 and 408 $cm^{-1}$. Moreover, MoS$_2$ also showed a weak 2LA(M) mode, which occurs from 453 to 465 cm$^{-1}$, ascribed to a two LA phonons process at the M point.[33a] When both hBN and MoS$_2$ were combined with EC, bands of the compound arouse proportionally to its concentration, similar to the observed in GNP/EC.

Therefore, from all the stated about FTIR and Raman analyses, it was possible to confirm the composition of the nanomaterials/EC/toluene suspensions.

### 3.2. Shear Rheology of 2D Nanoparticle Suspensions

To study 2D nanoparticles suspensions rheologically, it is primarily necessary to study the shear rheology of the carrier fluid. **Figure 6** shows steady-shear viscosity (η) curves for different polymer solutions of ethyl cellulose with toluene. The concentrations used are 0% w/v (pure toluene), 2.5% w/v and 5% w/v. When the carrier fluid is toluene the steady-shear viscosity is constant and independently of the shear rate applied, i.e., toluene behaves like as a Newtonian fluid and the zero-shear viscosity is equal $0.54 \pm 0.03$ mPa.s which is in accordance with the theoretical value published in literature.[34] It can be observed that the addition of ethyl cellulose increases the viscosity and changes the rheological behavior, promoting the shear-thinning behavior (**Figure 6**). For the solution of 2.5% w/v of ethyl cellulose, the shear-thinning behavior appears when the shear rate applied is higher than $10^4$ s$^{-1}$. Although, the shear-thinning is clearly observed for shear rates above $10^3$ s$^{-1}$ when the polymer solution has a concentration of 5% w/v of ethyl cellulose.

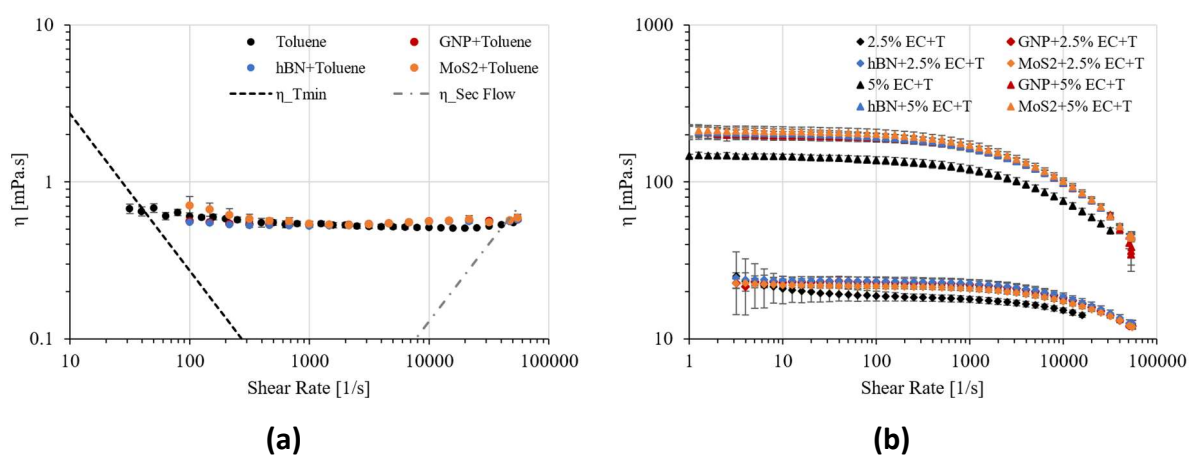

**Figure 6.** Steady-shear viscosity curves for all fluids studied in this work: **(a)** suspensions have toluene as carrier fluid and **(b)** suspensions have polymeric solution as carrier fluid.

**Figure 6a** shows steady-shear viscosity curves for all suspensions studied in this work when the carrier fluid is pure toluene. The presence of particles does not affect the shear viscosity when compared to the steady-shear viscosity curve of the carrier fluid. Moreover, the concentration of particles is about 230 ppm of weight by weight which does not change the characteristic properties of the carrier fluid. According to Giudice et al.[35], when the particle concentration is below than the critical concentration, the steady-shear viscosity of the suspension is similar to the steady-shear viscosity of the carrier fluid because no interactions between particles are present and, thus, isolated particles align immediately along the flow direction. This phenomenon was also observed by Thapliyal et al.[36] who found that the viscosity values of suspensions were very similar to the viscosity of the carrier fluid for the same shear rate applied when the concentration of cupper nanoparticles is equal to 0.2% w/w. As stated by Einstein's equation (**Equation 1**), it would be expected that the viscosity of the suspension ($\eta$) would be equal to the viscosity of the carrier fluid ($\eta_s$), because the volume fraction ($\phi$) of particles in these suspensions are in order of $10^{-5}$ in such a way that the second term of the Einstein's equation is negligible.

$$\eta = \eta_S(1 + 2.5\phi) \tag{1}$$

**Figure 6b** shows the steady-shear viscosity curves for suspensions which the carrier fluid is a polymeric solution with a concentration of 2.5% w/v of ethyl cellulose. The presence of particles increases the shear viscosity independently of the shear rate applied. However, the large standard deviation of each experimental point in the carrier fluid below 30 s$^{-1}$ did not allow to distinguish the presence of particles. Above 30 s$^{-1}$ it can be observed a clear distinction of the steady-shear viscosity curves and distinguish between two well defined regions for all suspensions:[37] The first region represents the plateau value of zero-shear viscosity for shear rates below $10^3$ s$^{-1}$, where the shear viscosity shows almost independence of shear rate applied; followed by the shear-thinning region, where the shear viscosity decreases when the shear rate increases.

When the carrier fluid is a polymer solution with a concentration of 5% w/v of ethyl cellulose, the presence of particles also induced an increment of shear viscosity for all shear rates applied (**Figure 6b**). Again, the same two regions can be observed, i.e., a plateau for the shear rate range at low shear rates followed by a shear-thinning region. Comparing the effect of the presence of the 2D nanoparticles in both viscoelastic solutions, the larger the concentration of EC in the system, the larger the increase of viscosity (

**Table 3**), but also the onset of shear thinning behavior seems to be shifted to higher shear rates.

**Table 3.** Zero-shear viscosity for all fluids studied in this work.

| | Zero-Shear Viscosity [*mPa.s*] | | | |
|---|---|---|---|---|
| Carrier Fluid \ Particles | **Without Particles** | **GNP** | **hBN** | **MoS$_2$** |
| **Toluene** | 0.54 ± 0.03 | 0.55 ± 0.02 | 0.55 ± 0.02 | 0.56 ± 0.02 |
| **2.5% EC + T** | 18.6 ± 0.4 | 23.0 ± 0.7 | 23.4 ± 0.2 | 21.7 ± 0.3 |
| **5% EC + T** | 145 ± 2 | 195 ± 2 | 206 ± 4 | 209 ± 2 |

**Table 3** allows to distinguish the effect of the different 2D nanoparticles on the zero-shear viscosity of the suspensions when the carrier fluid is viscoelastic: at a concentration of 2.5%w/w of EC, GNP and hBN exhibit the same value of zero-shear viscosity, which was slightly higher than in the MoS$_2$ suspension; at 5%w/w of EC, it is possible to observe that different 2D nanoparticles lead to sensibly different values, being minimal for GNP and maximal for MoS$_2$. The mechanics of suspended particles in non-Newtonian fluids is not nearly as understood as the mechanics of suspended particles in Newtonian fluid, even less in the case of 2D nanoparticles, mainly due to 'the polymers in solution act as "other particles" in a very similar sense to the "other particles" in a non-dilute particle suspension', as stated by Shaqfeh.[38] Recently, Einarsson *et al.*[39] introduced the first correction to the effective viscosity for an infinitely dilute suspension of spherical particles in a viscoelastic. However, this correction also fails in predicting the zero-shear viscosity values shown in

**Table 3**, as in 2D nanoparticles the orientation distribution also adds complexity in the system.[40]

Considering the size of the particles from DLS data (section 3.1.1), despite GNP exhibiting the largest D$_{90}$ values, it showed the largest size polydispersity, whereas MoS$_2$ was the opposite. Thinking of spherical particles, one would expect larger viscosities for those suspensions with larger particles sizes; however, in systems laden with 2D nanoparticles, the flow-induced orientation plays a dominant role. According to Reddy *et al.*[41], particle size polydispersity affects the average rotational diffusion coefficient. Considering the data obtained here, it seems that particles have a large particle size polydispersity can be packed in a better way than a suspension of particles of the same size or with a narrow particle size distribution. When the particle size polydispersity is larger, more free space is available for individual particles to move around when a flow is imposed to the suspension.[42] This leads to a lower viscosity of the ink. Moreover, flow-induced alignment of the 2D nanoparticles would be lead to an increase in the effective percolation threshold.[43]

## 3.3. Extensional Rheology of 2D Nanoparticle Suspensions

Extensional rheology is of paramount importance to the processing of viscoelastic 2D suspensions by any printing technique.[6] Moreover, in order to model and predict the fluid-flow behavior under actual printing conditions one needs to complement the shear rheometry with extensional rheometry.[44] Nevertheless, very few experimental works have been reported in the literature for spherical particles dispersed in weakly viscoelastic liquids[19a] and even less for 2D nanoparticles.[5] Einarsson *et al.*[39] developed an expression for spherical particle supensions in weakly viscoelastic fluids able to predict that the extensional viscosity of the suspension thickens even more drastically than the shear viscosity with increasing extension rates, but nothing has been developed yet for 2D nanoparticle laden viscoelastic fluids.

Before analyzing the behavior of the suspensions when subjected to uniaxial strain, it is important to characterize the extensional rheology of their carrier fluids. Toluene is an inelastic fluid which dominant material parameter in CaBER tests is the Ohnesorge number ($Oh$). The $Oh$ number is calculated according to **Equation 2** and it compares the time scales of viscosity-controlled breakup and inertia-controlled breakup.[22]

$$Oh = \frac{\eta}{\sqrt{\rho \sigma R}} \quad (2)$$

Being $\eta, \sigma$ and $\rho$ the viscosity, the surface tension, and the density of the fluid, respectively, whereas $R$ is the minimum radius of the liquid bridge. Once the Oh number is always less than 0.2077 for pure toluene.[22] This means that the decrease of the filament radius over time is controlled by the balance of viscous, inertial, and capillary forces and the minimum radius follows the following equation:[45]

$$R_m(t) = \frac{2X-1}{6} \frac{\sigma}{\eta} (t - t_b) \quad (3)$$

where X is a constant that is determined experimentally at the last points of $R_m(t)$ before the filament breaks. In this case, the value of $X$ for toluene is close to the value determined by Eggers (X = 0.5912)(**Table 4**).[46]

**Table 4.** Density (ρ), surface tension (σ), Ohnesorge number (Oh, Eq. 2) and X (Eq. 3) for all toluene suspensions at a temperature of 20 °C.

| Fluid | ρ [g/cm$^3$] | σ [mN/m] | Oh [-] | X [-] |
|---|---|---|---|---|
| Toluene | 0.865 | 27.38 ± 0.01 | 0.0027 | 0.5125 |
| GNP + Toluene | 0.859 | 24.84 ± 0.04 | 0.0029 | 0.5188 |
| hBN + Toluene | 0.864 | 27.38 ± 0.02 | 0.0025 | 0.5190 |
| MoS$_2$ + Toluene | 0.859 | 27.07 ± 0.02 | 0.0026 | 0.5191 |
| 2.5% w/v EC + Tol. | 0.871 | 27.80 ± 0.01 | | |
| GNP + 2.5% w/v EC + Tol. | 0.874 | 25.30 ± 0.03 | | |
| hBN + 2.5% w/v EC + Tol. | 0.875 | 27.23 ± 0.02 | | |
| MoS$_2$ + 2.5% w/v EC + Tol. | 0.874 | 27.96 ± 0.03 | > 0.2077 | -- |
| 5% w/v EC + Tol. | 0.908 | 26.44 ± 0.25 | | |
| GNP + 5% w/v EC + Tol. | 0.888 | 26.97 ± 0.06 | | |
| hBN + 5% w/v EC + Tol. | 0.879 | 25.73 ± 0.05 | | |
| MoS$_2$ + 5% w/v EC + Tol. | 0.882 | 27.97 ± 0.05 | | |

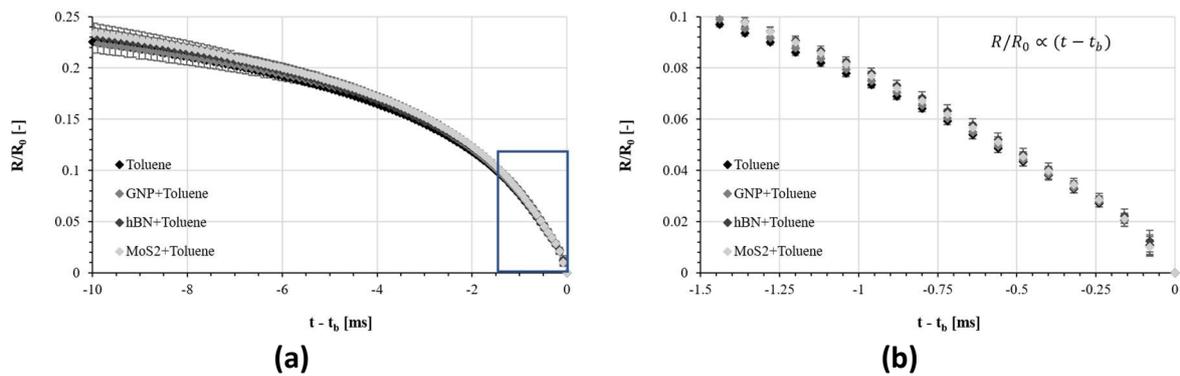

**Figure 7. (a)** Time evolution of the minimum filament radius for suspensions where toluene is the carrier fluid, **(b)** a scale zoom of the thinning process inserted into the blue rectangle.

The time evolution of the minimum filament radius curves for suspensions based on pure toluene are presented in **Figure 7a**. It can be observed in the zoomed in curve (**Figure 7b**) that, at the last stages before breaking up, the time evolution of the minimum radius follows the same linear relationship, independently of the particle dispersed in the system. As stated by Mathues *et al.*[47], this region is characterized by the absence of particles during the thinning process and the time evolution of the minimum filament radius is described by the **Equation 3**. Based on the X's values present in **Table 4,** all studied suspensions have practically the same value of X and it is close to the Eggers' solution which means that the thinning process is controlled by the balance of viscous, inertial, and capillary forces.

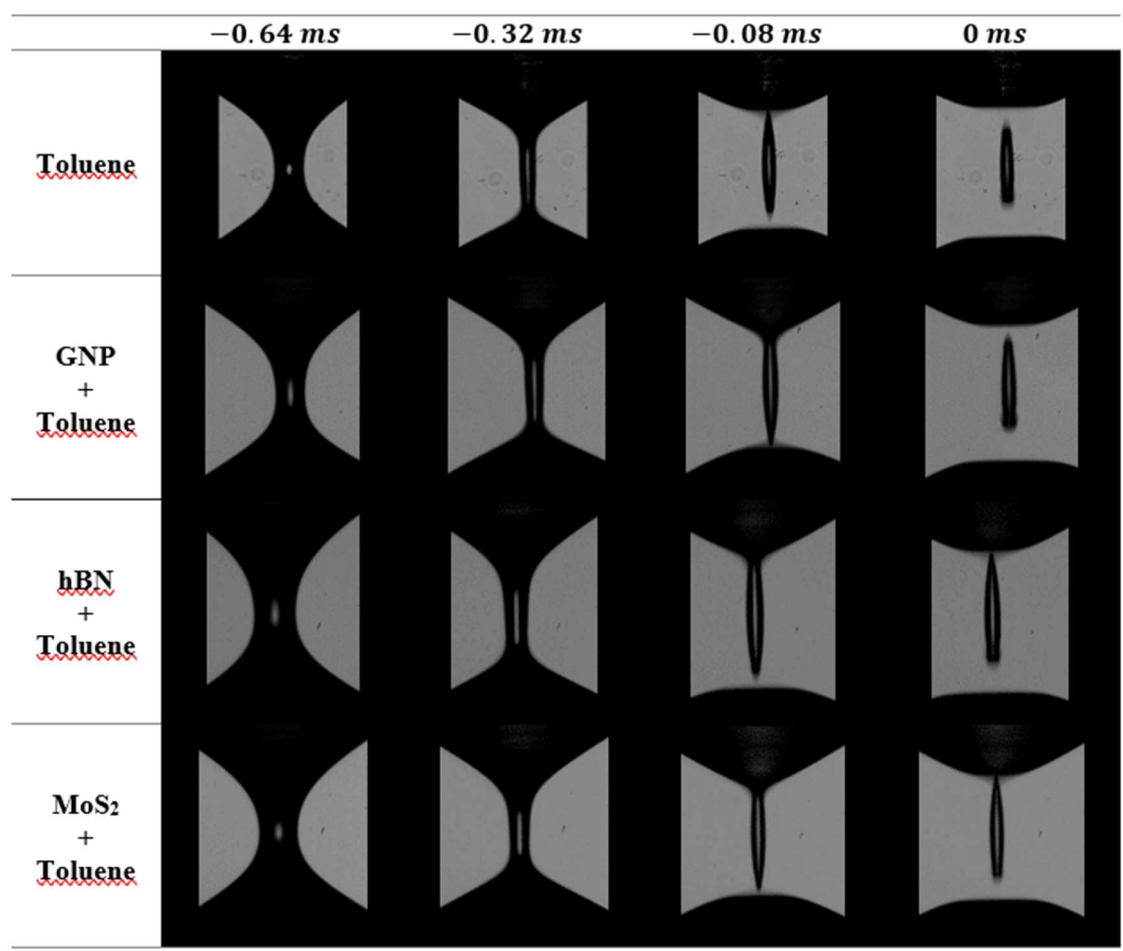

**Figure 8.** Last 0.64 ms of each filament process for suspensions without ethyl cellulose.

**Figure 8** shows the evolution of the filament thinning process for suspensions in pure toluene. It is observed that pinch-off is the preferred filament breaking model for these suspensions. As reported by Zhao *et al.*[48], the pinch-off is frequent when the suspension homogeneity condition is not respected, i.e., when there is an inhomogeneous distribution of particles in the fluid. When the minimum filament diameter approaches the particle size, the suspension cannot longer be considered homogeneous due to the migration of the particles outside of the minimum diameter.[47, 49] After that, the diffusion of particles is not possible to occur, so only the solvent controls the filament process in the pinch-off zone. As suspensions behave similarly to a Newtonian fluid, it would be expected that the breakage of the filament at the pinch-off ends would occur simultaneously. However, this does not happen because the inertial effects present in the fluid influence the filament thinning process.

**Figure 9** shows the evolution of the extensional viscosity ($\eta_E$) as a function of the extensional rate ($\dot{\varepsilon}$) for all suspensions which toluene is the carrier fluid. These plots were determined from the time evolution of the minimum diameter based on the following equations, respectively:[45]

$$\dot{\varepsilon} = -\frac{2}{R_m(t)} \cdot \frac{dR_m(t)}{dt} \tag{4}$$

$$\eta_E = \frac{\sigma}{R_m(t)} \cdot \frac{1}{\dot{\varepsilon}} \tag{5}$$

As the analysis of the filament necking dynamics in the CaBER is made in the region corresponding to the minimum diameter, far from the plates, the conclusions made can be directly extrapolated for the study of the dynamics of the droplet generation that will be analyzed later on.[50] In contrast to the steady shear viscosity measurements, the extensional flow by CaBER device allows the analysis of the transient extensional viscosity; consequently, it is possible to observe the effects of the orientation and migration of the nanoparticles on it.

The apparent extensional viscosity of toluene decreases monotonically as the radius of the filament decreases and that is inversely proportional to the increase in the extension rate. The Newtonian fluid condition is respected for high $\dot{\varepsilon}$ rates where the Trouton's rule is followed for uniaxial extensional flow, corresponding to the final linear decay of the minimum diameter with time before filament breaks up (**Figure 7b**). The GNP suspension also exhibit a monotonic decrease in extensional viscosity with the increasing extension rate at early stage; then, around 3000 s$^{-1}$, it rises up, reaches a maximum and then continue decreasing with a final value respecting the Trouton ratio (**Figure 9)**. The rise in the viscosity might be due to the rotation of the particles when the size of the filament is reduced[51] and the combination of having a large $D_{90}$ value with the confinement may difficult the alignment of the particles in the direction of the flow, which results in an extra internal resistance in the flow;[47] further increasing the extension rate lead to an extensional viscosity satisfying the Trouton ratio. In the case of hBN, the extensional viscosity decreases with the increase of the extension rate, and when reaches 3000 s$^{-1}$ the extensional viscosity becomes constant (**Figure 9)**, again probably due to the rotation of the particles (smaller $D_{90}$ than GNP) inside the thin filament; later on, the viscosity reaches the Trouton ratio. Finally, the MoS$_2$, the one having smallest $D_{90}$ value, exhibit a monotonic decrease of the extensional viscosity, with two changes in the slope, one around 1000s-1, for the rotation of the particles and a final one leading to the Trouton ratio (**Figure 9**). It is important to highlight the accomplishment of Trouton ratio at the end of the experiment, independently of the type of 2D nanoparticle dispersed in Toluene. This result might have been due to the particle migration, as reported for spherical and micro-sized particle suspensions[47], but it does not seem plausible because that would imply aggregates of the order of 20 μm; it may be more reasonable to hypothesize with the remanent particles being perfectly aligned with the flow direction.

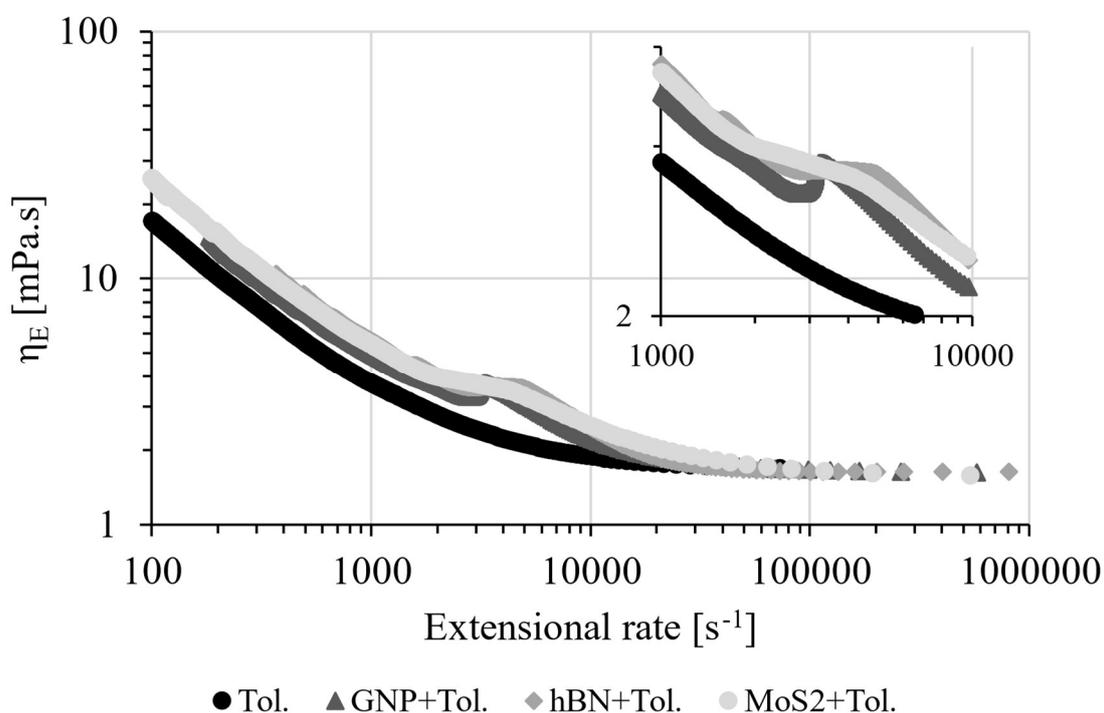

**Figure 9.** The extensional viscosity ($\eta_E$) as function of extensional rate for all suspensions in which toluene is the carrier fluid.

Regarding the suspensions where the carrier fluid is a polymeric solution with a concentration of 2.5% w/v of ethyl cellulose, **Figure 10** shows the time evolution of the minimum radius normalized by the radius of the end-plates. The homogeneity condition of the suspension is respected during the first 4 ms of the plot. Between -6 and 0.4 ms, it can be observed that the type of particles dispersed in the medium changes the time evolution of the minimum filament radius. Immediately before the filament breaks, curves overlap again, which may show that either the thinning process occurs in regions without particles, i.e., the thinning process is controlled only by the carrier fluid[47, 49], or the particles are completely aligned with the flow direction. For the time interval squared in **Figure 10**, the Ohnesorge number is always greater than 0.2077, which shows that the inertia of the fluid does not contribute to the balance of forces that sustain the filament, that is, only elastic and capillary forces are responsible for the filament thinning process. Moreover, since the Bond number is less than 0.2, the gravity effects can be also neglected.[22] In this region of the curve, the elastic fluid condition is respected and the relaxation time ($\lambda$) is calculated through the time derivative of the following equation:[45]

$$\frac{R_{min}}{R_0} = \left(\frac{GR_0}{2\sigma}\right)^{1/3} \exp\left(-\frac{t}{3\lambda}\right) \qquad (4)$$

where $G$ is the shear modulus. The relaxation times for each suspension are presented in **Figure 11**. Based on the data available in **Figure 10** and **Figure 11**, it is possible to conclude that suspensions with lower relaxation times promote a faster thinning process compared to the suspensions with higher relaxation time. So, it would be expected that the lifetime of the filament would be longer for the carrier fluid and suspension with hBN particles and lower for suspensions with GNP and $MoS_2$ particles.

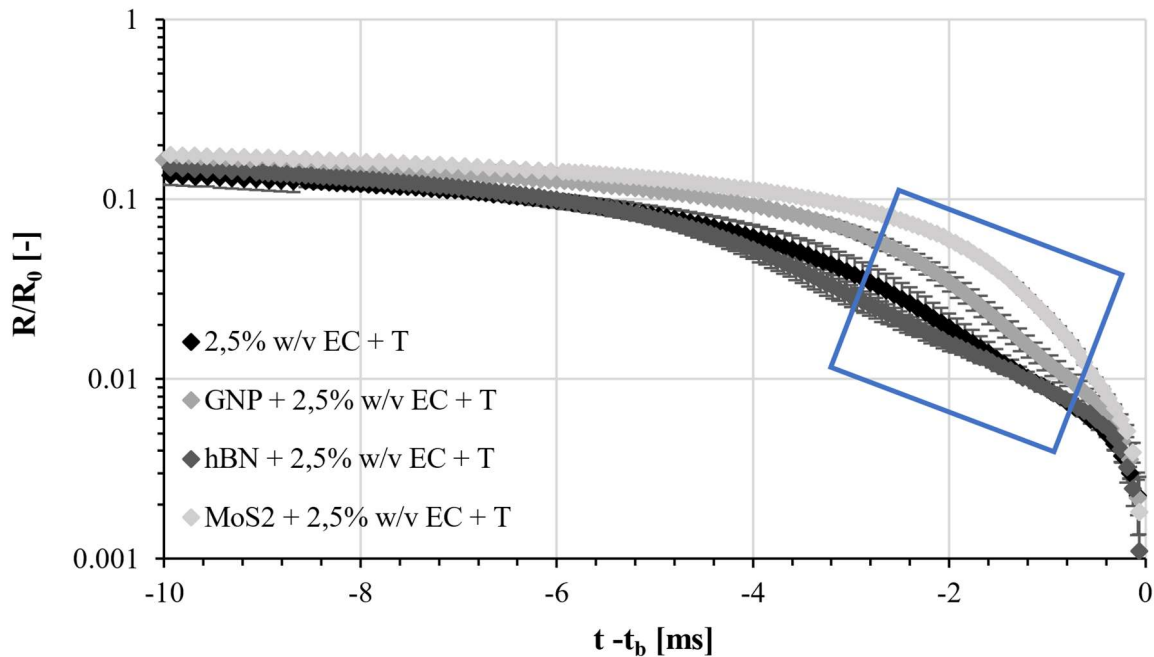

**Figure 10.** Time evolution of the minimum radius for suspensions has polymer solution 2.5% w/v of ethyl cellulose as a dispersant fluid.

The presence of nanoparticles in the polymer solution reduces the relaxation time compared to the relaxation time obtained for the polymer solution (**Figure 11**). This relaxation time reduction is also reported by Han and Kim for rod nanoparticles dispersed in a polymeric solution.[52] The authors argued that the particles are fully aligned along the extension direction and the flow near the particle becomes a shear flow to match the no slip boundary condition at the particle surface and hence the polymers close to the particle are tumbled and less stretched. However, for the hBN suspension the relaxation time is very close to the relaxation time obtained for the polymeric solution. This leads us to suspect that the slipping between the hBN particle and ethyl cellulose exists, which reduces the local shear flows near the particle and the polymers can be fully stretched. Although the three different nanomaterials used in this work share the same morphology (sheets), the slipping effect between hBN and ethyl cellulose is more pronounced than the slipping effect observed for GNP or MoS$_2$ with ethyl cellulose.

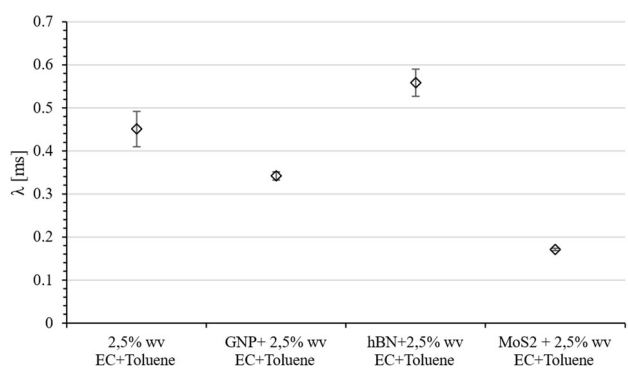

**Figure 11.** Relaxation time for suspensions with 2.5% w/v of ethyl cellulose.

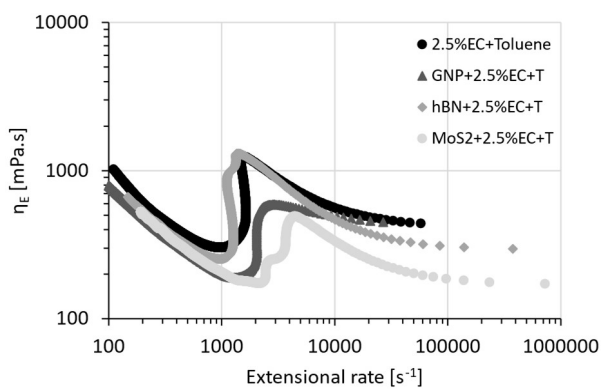

**Figure 12.** The extensional viscosity ($\eta_E$) as function of extensional rate for all suspensions in which 2.5% w/v of ethyl cellulose dissolved in toluene is the carrier fluid.

The reduction of relaxation time is also accompanied by a decrease in the extensional viscosity when nanoparticles are dispersed in the viscoelastic fluid (**Figure 12**). According to Han and Kim[52], when a pure polymer solution is subjected to the extensional flow, polymer chains are extended increasing the extensional viscosity until the polymer the finite extensibility. Consequently, **Figure 12** shows the narrow extensional rate where the extensional viscosity occurs, i.e., the strain hardening of the filament for the polymer solution. When nanoparticles are added to the fluid, the extensional viscosity decreases since the purely extensional flow condition is replaced by a combination of local shear flow condition near the particles and extensional flow condition far away the particles.[52] The local shear flow is generated by the hydrodynamic effects of the 2D nanoparticles. Although the hBN suspension has a similar relaxation time of the suspending fluid, **Figure 12** shows that the extensional viscosity of hBN suspension is lower than the extensional viscosity of the suspending fluid, which means that the local shear flow is present even though the slipping effect.

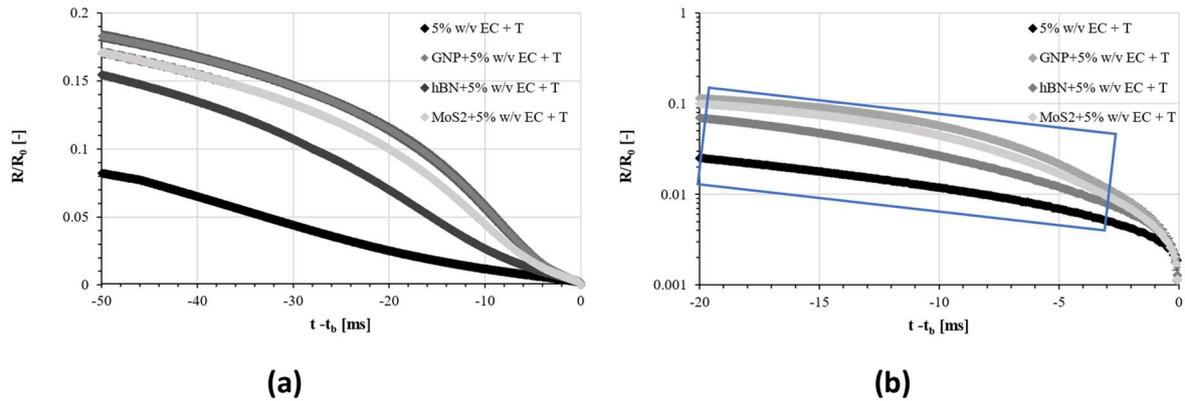

**(a)** **(b)**

**Figure 13. (a)** Time evolution of the minimum filament radius for all suspensions has a polymeric solution of 5% w/v of ethyl cellulose as a carrier fluid. **(b)** Magnification of the minimum filament radius curves for the last 5 ms of the filament lifetime before the breakage.

When the carrier fluid is a polymeric solution with a concentration of 5% w/v of ethyl cellulose dissolved in toluene, **Figure 13** shows a clear distinction about the thinning process when the particles are dispersed in the medium. However, the suspensions follow the same tendency. For the time scale available in **Figure 13a**, there is a three slope changes. The first slope corresponds to the concentration fluctuation regime, where the homogeneity condition for suspensions is not respected and will start to appear regions rich in particles or poor in particles along the filament.[47]. After that, an exponential decay of the filament radius is observed, and the polymer chains are stretched in this phase. When this happens, it occurs a strain hardening of the filament and the thinning process is controlled by a balance between the elastic stress and capillary pressure.[47, 52] Finally, the fluids show a decelerated regime where the properties of the solvent (toluene) control the thinning process.[47] In this phase, it is expected that the polymer chains are fully stretched and migrate to the ends of the liquid bridge. During the thinning process, the effects of inertia and gravity are neglected since the Ohnesorge number is higher than 0.2077 and Bond number is lower than 0.2 for all fluids.

When the carrier fluid is 5% w/v of ethyl cellulose dissolved in toluene, the suspensions behave like as an elastic fluid since there is a linear relationship between the filament radius and time in a semi-logarithmic graph present in **Figure 13b**. The relaxation time is calculated through the **Equation 4** for all experimental values inside the blue box present in **Figure 13b**. From the **Figure 14**, it is possible to observe that the addition of particles reduces the relaxation time of the suspending fluid, and these results agree with the data available in **Figure 13** since the higher is the slope of the experimental curve, the shorter is the relaxation time of the fluid. Further, these results agree with the tendency observed for the fluids with lower polymer concentration and the explanation stated for the fluids with 2.5% w/v of ethyl cellulose is also valid for this case.

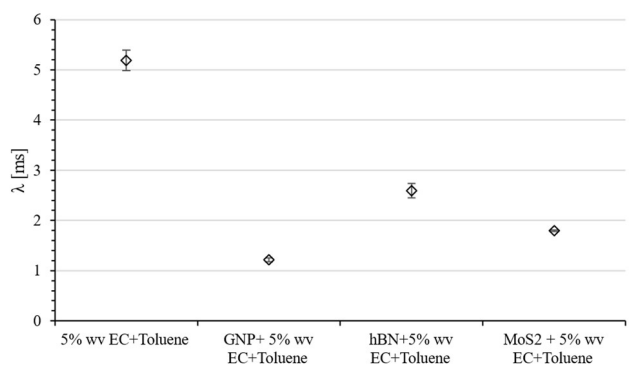

**Figure 14.** The relaxation time ($\lambda$) for all suspensions have 5% w/v of ethyl cellulose dissolved in toluene as a carrier fluid.

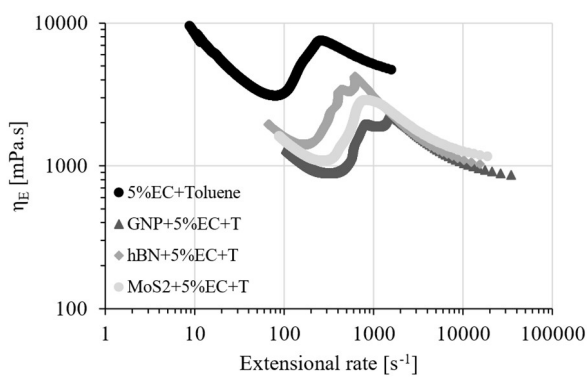

**Figure 15.** The extensional viscosity ($\eta_E$) as function of extensional rate for all suspensions in which 5% w/v of ethyl cellulose dissolved in toluene is the carrier fluid.

**Figure 15** shows the extensional viscosity curves for all suspensions which the carrier fluid is a polymeric solution with a concentration of 5% w/v of ethyl cellulose. For this carrier fluid, the behavior of the $\eta_E$ curves are similar to the ones obtained for the polymeric solution with 2.5% w/v of ethyl cellulose. In **Figure 15** it is observed that the polymer chains are stretched for thick extensional rate range even though the particles decrease the extensional viscosity values. The reduction of the extensional viscosity is due to the presence of local shear flow near the particles which hinders the polymer to stretch. As stated by Han and Kim[52], the purely extensional flow condition is no longer respected, on the other hand, the thinning process is controlled by a combination of shear and extensional flows. According to Jain *et al.*[40], there are two contributions of the particles that can affect the average stress/viscosity of suspensions: the stresslet and the particle induced fluid stress (PIFS). The PIFS occurs every time a particle in the fluid creates polymer stretch, i.e., nonlinear elastic response. According to Zhang and Shaqfeh[53] for highly shear-thinning fluids, the suspensions exhibit greater shear-thinning of the viscosity than the suspending fluid itself, which agrees with the results shown in **Figure 6**, **Figure *12*** and **Figure 15**. As stated by them[53], this characteristic of the fluid occurs due to the inability of the per particle viscosity of PIFS to overcome the shear-thinning per particle viscosity from the stresslet even though the shear-thickening profile of the per particle viscosity of PIFS. This inability of the PIFS is due to the presence of spiral streamlines near the surface particles which flows away a small volume of suspending fluid in a short time interval which affects the polymer stretch, i.e., the per particle viscosity of PIFS.

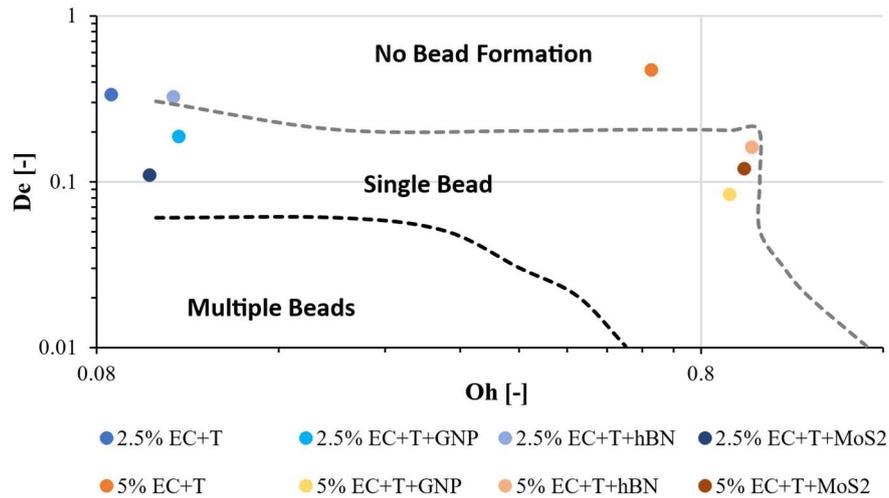

**Figure 16.** Phase diagram depicting the regions showing different BOAS morphologies in the Deborah number (De) and Ohnesorge number (Oh) space adapted from Bhat et al.[54]

The beads-on-a-string (BOAS) structures consist of single or multiple spherical fluid droplets interconnected by slender threads. This phenomenon appears when an extensional strain rate is imposed on a viscoelastic fluid. According to Bhat *et al.*[54], BOAS structure is created whenever there is a balance between two dimensionless numbers: Deborah (*De*) and Ohnesorge numbers. These beads can subsequently become undesirable satellite droplets, by increasing the polydispersity of the resultant droplet population.[55] Despite the phase diagram depicted in **Figure 16**, it was conceived that for polymeric solutions without the presence of particles, this theoretical prediction agrees with the experimental present in **Figure 17** for suspension with 2.5% w/v of ethyl cellulose. This happens because the homogeneity condition is not respected anymore when the diameter of the liquid bridge is equal or lower to the diameter of aggregates formed by a mixture of particles and polymer.[47-49] Thereafter, the diffusion of particles and polymer is suppressed and the solvent controls the dynamic of the thinning process.

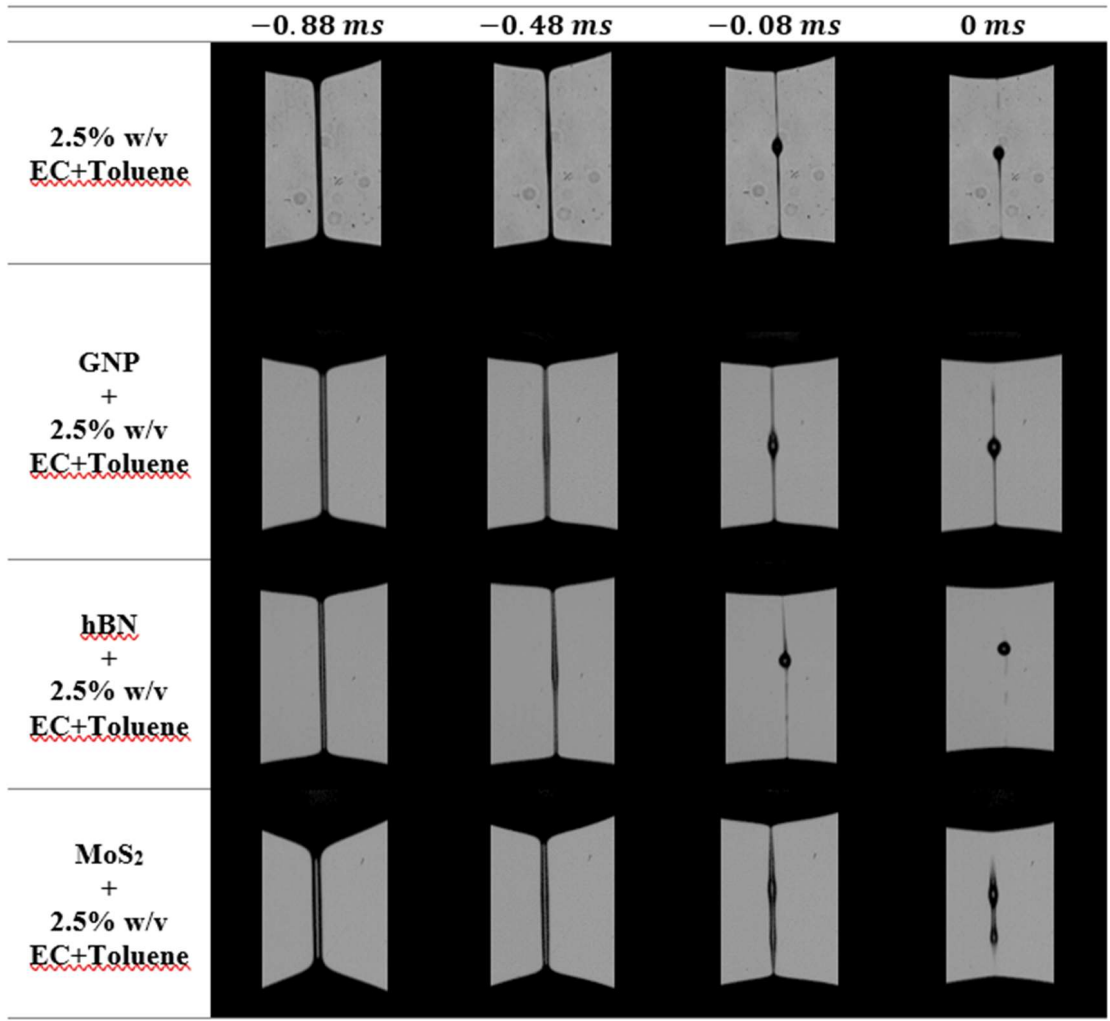

**Figure 17.** Last 0.88 ms of each filament process for suspensions with 2.5% w/v of ethyl cellulose.

**Figure 18** shows the evolution of the filament thinning process for suspensions with 5% w/v of ethyl cellulose. It can be observed that the absence of bead formation in the liquid bridge. These experimental results are inconsistent with the prediction shown in **Figure 16**, although the limits proposed by Bhat *et al.*[54] are valid to define the BOAS formation region for microparticles dispersed in a polymeric solution.[19a] Based on the experimental results, it can hypothetically be inferred that the particle shape and the polydispersity of particles can reduce the critical Deborah number at which there is a transition from a liquid bridge with a single bead to a liquid bridge without beads. This contradictory result asks for further theoretical, computational and experimental research, which is out of the scope of the current work.

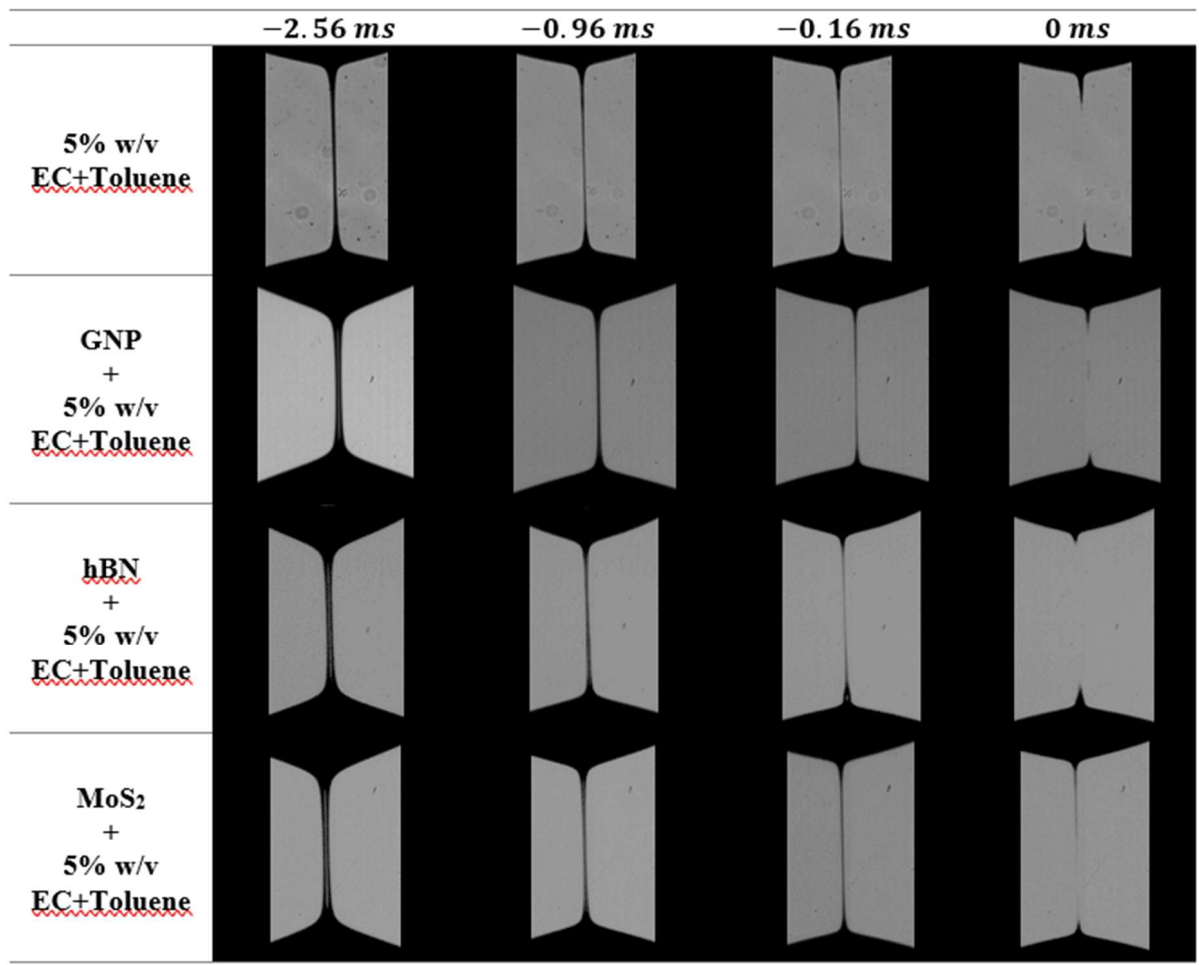

**Figure 18.** Last 2.56 ms of each filament thinning process for suspensions with 5% w/v of ethyl cellulose.

### 3.4. Droplet size in dripping regime

In this section, for the sake of simplicity, only the Newtonian carrier fluid (pure Toluene) and one viscoelastic carrier fluid (Toluene+2.5% w/v EC) will be considered. The size of the droplets generated in dripping regime will be correlated with the rheological information discussed above. **Figure 19** shows the sequence of images during taken for the same nozzle and a specific flow rate.

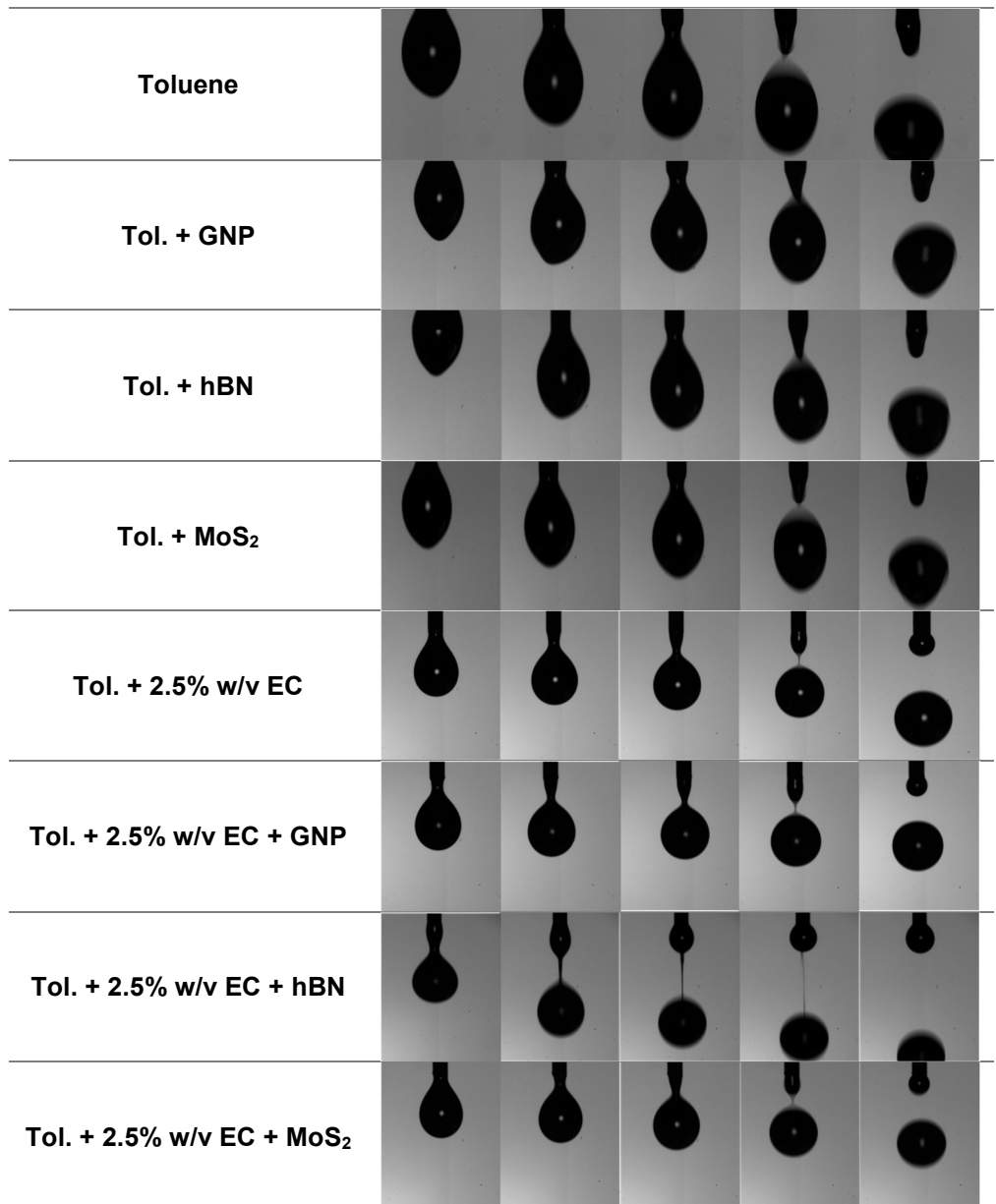

**Figure 19**. Dispensing in the dripping regime of the different working fluids from a nozzle with an inner diameter of 0.41 mm with a flow rate of 50 μL/s. Note: the optics zoom used for the first four fluids is higher than the one used for the later ones.

**Figure 19** shows the photographic record of droplet formation for all fluids studied here for an internal nozzle diameter of 0.41 mm and a flow rate of 50 μL/s. For fluids without polymer, it is observed that the fluid wets the outer surface of the nozzle and part of it rises along the outer surface due to capillary effects. This phenomenon is problematic for injectors in printers, where residual inks on the nozzle surface can change the droplet trajectory and droplet size which affect the quality of the printing process.[56] According to Sedighi *et al*.[57], the capillary-driven rise is affected by the liquid properties, nozzle size and flow rates. In nondimensional point of view, this phenomenon is affected by Ohnesorge ($Oh = \eta/\sqrt{D_{in}\rho g}$), Weber ($We = \rho U^2 D_{in}/\sigma$) and Bond ($Bo = \rho g D_{in}^2/\sigma$) numbers. Further, the outer surface of the nozzle is only wetted in dripping regime and the capillary-driven rise vanishes in jetting regime for Newtonian fluids. According to the authors[57], there is a critical Weber number that limits the capillary-driven rise for a fixed nozzle diameter and fluid's properties. **Table 5** shows that the critical Weber number is almost constant for a fixed nozzle diameter when the suspensions do not have ethyl cellulose in their composition. This happens because the fluids' properties are practically the same (

**Table 3 and Table 4**) for dilute particles concentration as discussed in **section 3.2.** Further, the critical We number increases when the nozzle diameter decreases. The addition of polymer to the fluids changes their properties (

**Table 3 and Table 4**) and the capillary-driven rise effect disappears (**Figure 19**), i.e., the fluid just wets the tip of the nozzle, which agrees with the results obtained by Sedighi *et al.*[57] for viscous fluids. They found that the capillary-driven effects are present for a very low We numbers (We < $10^{-3}$) for fluids that have Oh > 0.1.

**Table 5.** Critical Weber number for each fluid and nozzle diameter.

|  | Toluene | | | |
| --- | --- | --- | --- | --- |
| $D_{noz_{in}}$ [mm] | Without Particles | GNP | hBN | $MoS_2$ |
| **1.36** | 0.386 | 0.424 | 0.387 | 0.389 |
| **0.61** | 1.124 | 1.235 | 1.127 | 0.819 |
| **0.41** | 1.333 | 1.464 | 1.336 | 1.129 |
| **0.20** | 1.993 | 2.190 | 1.998 | 1.852 |

Still in **Figure 19**, it is possible to observe how the inertia effects affect the droplet detachment for fluids without polymer. When polymer is present, it is possible to see a capillary thinning process similar to the one observed in CaBER experiments for viscoelastic fluids. According to Clasen *et al.*[14], for Newtonian fluids in the dripping regime, the volume of the drop determines the precision of dispensing at low-flow rates, which can be described by the Harkins-Brown relation (**Equation 5**):[58]

$$V_{drop} = f_{HB} \frac{2\pi\sigma R}{\rho g}, \qquad (5)$$

where $R$ is the inner radius of the nozzle orifice ($\frac{D_{in}}{2}$) in the case of a nonwetting material or the outer radius ($\frac{D_{out}}{2}$) in the case of a wetting material; the coefficient $f_{HB}$ accounting for the non-sphericity due to gravity follows **Equation 6**:[59]

$$f_{HB} = 0.6248 R^{-0.1352} \left(\frac{\sigma}{\rho g}\right)^{0.0676}, \tag{6}$$

for a range $0 < \frac{R}{V_{drop}^{1/3}} < 1$.

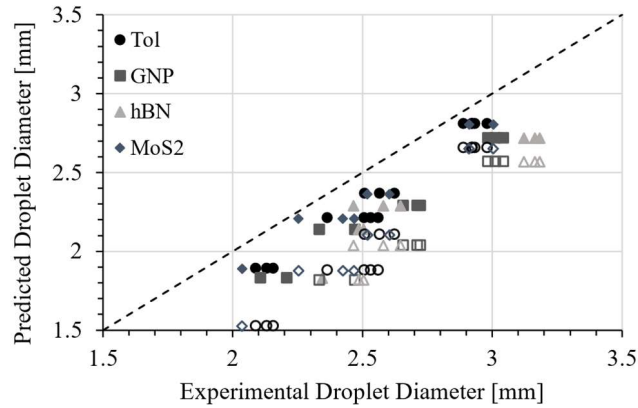

**Figure 20.** Comparison between the droplet diameter measured and the one predicted by **Equation 5**. Open symbols represent the nonwetting condition and filled symbols represent the wetting condition.

**Figure 20** shows a comparison between the experimental droplet size and the predicted one by **Equation 5**. Despite the intrinsic hydrophilic nature of stainless-steel, considering the outer radius in the calculation led to a better correlation between experiments and predicted sizes by **Equation 5 and 7**, due to inertial effects. It can be observed that **Equation 5** works better for pure Toluene, providing an average relative error between the experimental and the predicted droplet size of ~5%, than for the particulated fluids, which exhibit relative errors above 12%, 14% and 9% for GNP, hBN and MoS$_2$, respectively. The inertia controlled regime is corroborated by the presence of satellite droplets[14] potentially leading to a general unclean dispensing and inaccurate amounts of the dispensed fluid. Regarding the viscoelastic fluids, according to Clasen *et al.*[14], inviscid viscoelastic liquids are generally easy to dispense; however, viscoelastic solutions of high viscosity and with an important elastic response will represent serious difficulties due to nonlinear elastic stresses will slow the breaking dynamics dramatically. To the best of author's knowledge, there is no equation able to predict the droplet size for viscoelastic fluids.

Dimensional analysis is frequently used in Newtonian fluid mechanics to determine appropriate dimensionless groups for a particular problem, and it is also of common use for viscoelastic fluid flow problems.[60] Buckingham Π theorem[61] allows to reduce from nine variables, i.e. three geometric ($D_{in}, D_{out}, D_{drop}$), four related to material properties ($\sigma, \rho, \eta, \lambda$) and two associated with the flow process ($g$ and $U$), to 6 dimensionless groups:

Reynolds number: $Re = \frac{\rho D_{in} U}{\eta}$;

Weber number: $We = \frac{\rho U^2 D_{in}}{\sigma}$;

Froude number: $Fr = \frac{U^2}{g D_{in}}$, where $g$ = 9.81 m/s² is the gravity;

Weissenberg number: $Wi = \frac{\lambda U}{D_{in}}$;

And two geometric dimensionless numbers: $\pi_1 = \frac{D_{out}}{D_{in}}$ and $\pi_2 = \frac{D_{drop}}{D_{in}}$.

Dimensionless groups can be combined to reduce their number. Thus, the Ohnesorge number can be defined based on Weber and Reynolds numbers ($Oh = \frac{\sqrt{We}}{Re} = \frac{\eta}{\sqrt{D_{in} \rho g}}$) and the Bond number in terms of Froude and Weber ($Bo = \frac{We}{Fr} = \frac{\rho g D_{in}^2}{\sigma}$). **Table 6** shows the ranges for these four dimensionless numbers in our experiments, considering the different nozzles (**Table 1**), flow rates and fluid properties (**Table 2 and Table 3**). It can be observed that slight changes in $\eta, \sigma$ and $\rho$ result in similar ranges for $Bo$ in all the samples, and for $Oh$ for the Newtonian and the viscoelastic suspensions. As the Newtonian samples do not have relaxation time, the dimensional analysis implies one variable less and, consequently, the Weissenberg does not show up.

**Table 6**. Ranges of values for the different dimensionless parameters considered in this study.

| Fluid Sample | $\frac{D_{out}}{D_{in}}$ [-] | $Bo$ [-] | $Oh$ [-] | $Wi$ [-] |
|---|---|---|---|---|
| Toluene | 1.2 – 2.1 | 0.01 – 0.63 | 0.003 – 0.009 | - |
| Tol. + GNP | | | 0.003 – 0.009 | - |
| Tol. + hBN | | | 0.003 – 0.009 | - |
| Tol. + MoS₂ | | | 0.003 – 0.009 | - |
| Tol. + 2.5% w/v EC | | | 0.10 – 0.35 | 0.01 – 43.1 |
| Tol. + 2.5% w/v EC + GNP | | | 0.10 – 0.35 | 0.03 – 30.4 |
| Tol. + 2.5% w/v EC + hBN | | | 0.10 – 0.35 | 0.05 – 18.00 |
| Tol. + 2.5% w/v EC + MoS₂ | | | 0.10 – 0.35 | 0.01 – 5.44 |

**Figure 21** shows the normalized droplet size ($D_{drop}/D_{nozzle}$) as a function of three dimensionless parameters (Bo, Oh and We). In the diagram $D_{drop}/D_{nozzle}$ vs Bo number, the normalized droplet diameter decreases with increasing Bond number. Since Bo number depends exclusively on the density, surface tension of the fluid and nozzle diameter. According to **Table 4,** the density and surface tension of the fluids sharing the same carrier fluid (toluene or 2.5% w/v ethyl cellulose dissolved in toluene) are similar, which shows that the parameter $D_{drop}/D_{nozzle}$ decreases by diminishing the nozzle diameter.

However, it is not possible to observe a clear trend in the relationship between the applied flow rate and the droplet size, so it is necessary to draw the diagram $D_{drop}/D_{nozzle}$ vs We number. In this diagram (**Figure 21**) we can observe two trends. When the nozzle diameter is fixed and the flow rate is increased, the droplet diameter remains almost constant, only the number of drops ejected in the same time interval increases. When the flow rate (*Q*) is fixed, we observe that the normalized droplet diameter increases with decreasing nozzle diameter since $We = (16\rho Q^2)/(\pi^2 D_{nozzle}^3 \sigma)$.

However, these two diagrams only allow to characterize the influence of the droplet diameter with two fluid properties (density and surface tension) and 2 operating properties (nozzle diameter, flow rate). To study the influence of fluid rheology properties on droplet formation, the best dimensionless number to use is the Ohnesorge number. According to diagram $D_{drop}/D_{nozzle}$ vs Oh number, it can be seen that the normalized droplet diameter is not influenced when polymer is added to the fluid, i.e., when the viscosity increases. Nevertheless, this addition of polymer leads to inertial effects being damped/eliminated during droplet detachment. In this graph (**Figure 21**) it is also possible to observe that the droplet diameter is strongly influenced by the nozzle diameter, with $D_{drop}/D_{nozzle}$ being larger as the nozzle diameter decreases for the same fluid. This trend is also observed in the $D_{drop}/D_{nozzle}$ vs Bo and $D_{drop}/D_{nozzle}$ vs We diagrams, which reinforces the strong dependency that exists between droplet diameter and nozzle diameter.

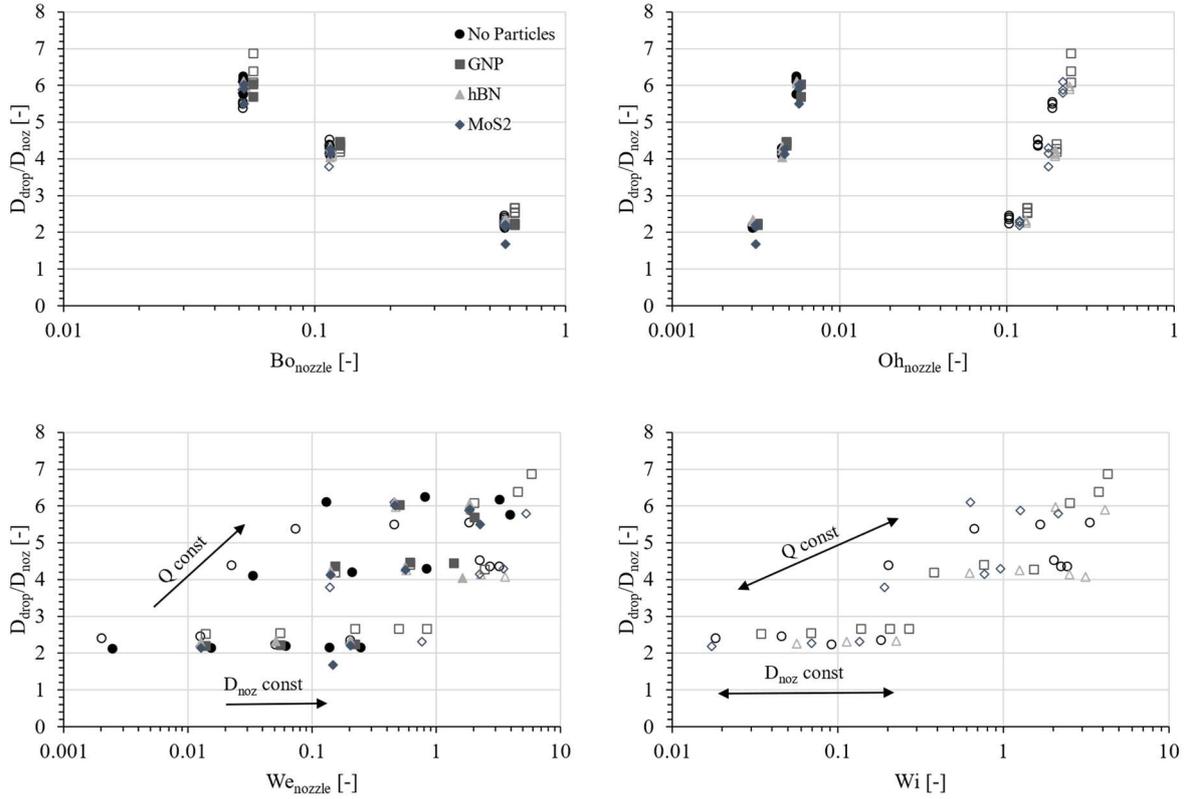

**Figure 21.** Maps for the normalized droplet size as a function of the different dimensionless numbers.

Relatively to the influence of Weissenberg number (Wi) on normalized droplet size (**Figure 21**), it is possible to observe the same tendencies observed on the $D_{drop}/D_{nozzle}$ vs We diagram. When the nozzle diameter is fixed and the flow rate increases, the normalized droplet diameter stays practically constant; however, the normalized droplet diameter increases when the flow rate is constant and the nozzle diameter decreases. According to Nooranidoost et al.[62], the viscoelasticity does not have a significant influence on the droplet size until a critical Wi number ($Wi_{cr}$) is reached. When the Wi number exceeds the critical value $Wi_{cr}$ ~0.6, the droplet size increases abruptly reaching another nearly constant value, which agrees with the results present on **Figure 21.** The $D_{drop}/D_{nozzle}$ vs Wi diagram shows three plateaus where it is possible to observe two distinct of $Wi_{cr}$ values. The first $Wi_{cr}$ value occurs at $Wi_{cr}$~0.2 and the second one occurs at $Wi_{cr}$~0.6, as reported by Nooranidoost et al.[62]. **Figure 22.** Influence of Weissenberg number (Wi) on normalized droplet size (Ddrop/Dnoz) for a flow rate of 25 μL/s. It can be clearly seen the three ranges limited by the critical Weissenberg values.

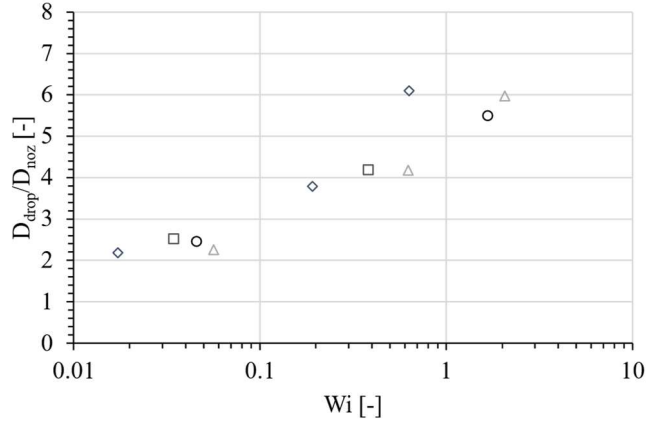

**Figure 22.** Influence of Weissenberg number (Wi) on normalized droplet size (D$_{drop}$/D$_{noz}$) for a flow rate of 25 µL/s.

Moreover, the Buckingham Π theorem allows to express one dimensionless group as a function of the other π-groups. Consequently, it is possible to predict the droplet size generated in terms of the relative dimensions of the nozzle ($\pi_1 = \frac{D_{out}}{D_{in}}$) and the properties of the fluid and the process ($Bo, Oh, Wi$). In many fluid mechanics problems, a generic power correlation typically works well (Equation 5):[63]

$$\frac{D_{drop}}{D_{in}} = A\pi_1^b Bo^c Oh^d Wi^e, \tag{7}$$

where $A, b, c, d,$ and $e$ are fitting constants. **Equation 7** is valid for fluids that have polymer in their composition. For Newtonian fluids, the last term of the above equation is ignored. To quantify the goodness of the correlation, a Mean Relative Error (MRE) will be evaluated, and it is defined as:

$$\text{MRE} = \sum_i \frac{\left|\left(\frac{D_{drop}}{D_{in}}\right)_{exp} - \left(\frac{D_{drop}}{D_{in}}\right)_{corr}\right|}{\left(\frac{D_{drop}}{D_{in}}\right)_{corr}} \tag{8}$$

where the subscript indicates if the corresponding values are experimental ones (exp) or evaluated by the correlation (corr), while the index i goes for all experimental cases. If the MRE values is smaller, **Equation 7** has a better fit to the experimental data.

**Table 5** summarizes the values of these constants for the different fluids. Regarding the MRE of the correlations, on the one hand, we could say that the correlation predicts reasonably well the values obtained experimentally. The dependence of the normalized droplet diameter on the ratio of the nozzle's diameters, the Bond, the Ohnesorge and Weissenberg numbers has been confirmed thanks to the proposed mathematical correlations. They can help to predict the droplet size for each suspension without having to carry out the setting up the experiments or without carrying numerical simulations. For any Newtonian particulate suspensions, **Equation 7** provides a better prediction than **Equation 5**, as it accounts for all the effects affecting the dispensing process. Additionally, it also predicts reasonably well the droplet size for the viscoelastic particulate suspensions.

Table 7. Correlation factors depending on the working fluid.

| Fluids\Constants | A [-] | b [-] | c [-] | d [-] | e [-] | RME [%] |
|---|---|---|---|---|---|---|
| Toluene | 2.8681 | 1.2193 | -0.2102 | 0.1063 | -- | 1.78 |
| Tol. + GNP | 3.5979 | -0.8984 | -0.5156 | 0.0946 | -- | 1.98 |
| Tol. + hBN | 2.4995 | 1.2177 | -0.1968 | 0.0697 | -- | 4.26 |
| Tol. + $MoS_2$ | 2.7005 | -0.6316 | -0.5192 | 0.0794 | -- | 5.93 |
| Tol. + 2.5% w/v EC | 0.0285 | 17.733 | 1.9153 | -1.0130 | -0.0238 | 9.95 |
| Tol. + 2.5% w/v EC + GNP | 2.5606 | 3.3481 | 0.1339 | 0.3515 | -0.0663 | 10.4 |
| Tol. + 2.5% w/v EC + hBN | 2.8346 | 0.8846 | -0.2112 | 0.2545 | -0.0116 | 6.08 |
| Tol. + 2.5% w/v EC + $MoS_2$ | 2.8693 | 1.3580 | -0.1117 | 0.2443 | 0.0161 | 3.93 |

## 4. Conclusions

This work presented a rheological characterization of suspensions of GNP, hBN, and $MoS_2$ nanoparticles in ethyl cellulose dissolved in toluene. The effect of concentration of ethyl cellulose was investigated. The presence of ethyl cellulose improved the stability of the dispersions, mainly due to its steric stabilization mechanism. The materials' characterization techniques showed that the particles have a nano-size even in the presence of ethyl cellulose, as seen through DLS technique.

The steady-shear measurements showed that toluene behaves as a Newtonian fluid, and the addition of ethyl cellulose changes the fluid behavior from Newtonian to a shear-thinning fluid. Further, the shear-thinning occurs for lower shear rates when the polymer concentration increases. When the 2D-nanopartiles were dispersed in toluene, the fluid behavior did not change, and the viscosity calculated from Einstein's equation agrees with the experimental one. Relatively to the polymer solution, the addition of 2D-nanoparticles also did not change the fluid behavior, but the shear viscosity increases when compared to the polymer solution.

Relatively to the extensional rheology experiments, the suspensions in which toluene is the carrier fluid, inertia is present during the filament thinning, but the pinch-off phenomenon is present in all suspensions. When polymer is added to the suspension, the inertia effects are negligible, and the filament radius decay can be fitted by the exponential equation. The relaxation times of suspension decrease when GNP and $MoS_2$ particles were added due to the slipping between particles and polymer. For a concentration of ethyl cellulose of 2.5% w/v, the beads-on-a-string phenomena are present for all suspensions. For higher concentration of polymer, the filament thins and breaks in the middle without the formation of BOAS.

Relatively to the droplet formation, the inertia effects promote the satellite drops formation in the formulation without EC; moreover, these inks wet the nozzle surface, which affects the droplet size. When ethyl cellulose is present in the ink, the two phenomena described above disappear and the droplet size can be well predicted through the generic power law taking into account the fluid's properties, flow rate, and nozzle's diameter.

**Supporting Information**

Supporting Information is available from the authors.

**Acknowledgements**

This work was also financially supported by UI/BD/150886/2021 (FCT), LA/P/0045/2020 (ALiCE), UIDB/00532/2020, and UIDP/00532/2020 (CEFT), funded by national funds through FCT/MCTES (PIDDAC); PTDC/EME-APL/30765/2017-POCI-01-0145-FEDER-030765 and the program Stimulus of Scientific Employment, Individual Support-2020.03203.CEECIND, funded by FEDER funds through COMPETE2020 – Programa Operacional Competitividade e Internacionalização (POCI) and by national funds (PIDDAC) through FCT/MCTES. The authors would also to acknowledge Conselho Nacional de Desenvolvimento Científico e Tecnológico (CNPq) for the grants 140241/2019-1 and 305109/2022-7; National Institute of Science and Technology for Rheology of Complex Materials Applied to Advanced Technologies (INCT-Rhe9) grant 406765/2022-7, Coordenação de Aperfeiçoamento de Pessoal de Nível Superior (CAPES) for the grant PRINT 88887.310339/2018-00. The authors also thank Graphenest for graciously provided graphene nanoplatelets, Prof. J. Ortega-Casanova for the help in the fitting of the dimensionless correlations, and Dra. L. Campo-Deaño for fruitful discussions.